\documentclass[aps, pra, showkeys, showpacs, reprint]{revtex4-1} % uses revtex4-1 package

\usepackage[usenames,dvipsnames]{color} %can use colors in the script
\usepackage{amsmath} % advanced maths in LaTeX
\usepackage{txfonts} % make fonts similar to PRL format fonts - important to include this AFTER amsmath as otherwise it will throw up an error since both packages define a command \iint
\usepackage{upgreek} %non-italicesed greek letters (e.g. mu)
\usepackage{todonotes} %can add in notes of things to do \todo{Add details...}
\usepackage{amssymb} %more math symbols, e.g. integer = \mathbb{Z}, complex = \mathbb{C}
\usepackage{braket} %bra-ket notation, e.g. $\bra{x}\ket{\psi}$
\usepackage{array} %allows you to do \eqnarray i.e. splitting functions across multiple lines

\usepackage{afterpage} % for adding blank pages

\usepackage{bbold} % mathbb

\usepackage{enumitem}

\usepackage[unicode]{hyperref} 
\hypersetup{
    unicode=true,                                           % non-Latin characters in Acrobat’s bookmarks
    a4paper=true,
    plainpages=false,
    pdffitwindow=true,                                      % page fit to window when opened
    pdftitle={Cooperative eigenmodes and scattering in 1D atomic arrays},                      % title
    pdfauthor={R J Bettles},                            % author
    pdfsubject={},                                          % subject of the document
    pdfkeywords={},% list of keywords
    colorlinks=true,                                        % false: boxed links; true: colored links
    linkcolor=blue,                                         % color of internal links
    citecolor=blue,                                         % color of links to bibliography
    filecolor=blue,                                         % color of file links
    urlcolor=blue                                           % color of external links
}
\usepackage[caption=false]{subfig} %subfigures, if don't have caption=false then the caption justification of figures becomes centered
\usepackage{graphicx} %to manage external pictures
\usepackage[countmax]{subfloat}
\usepackage{import} %import graphics from other paths
\usepackage{transparent} %use transparency in the graphics

\definecolor{MyGreen}{rgb}{0,0.5,0} % define particular colors

\newcommand{\polvec}{\ensuremath{\hat{\boldsymbol{\epsilon}}}} 

\graphicspath{ {../Figures/} }% add this path to the list of paths searched for graphics

\makeatletter
\newcommand{\colorcaption}[3][]{%
  \begingroup%
  \renewcommand{\@caption@fignum@sep}{ (color online). }%
  \caption[#1]{#2}
  \label{#3}% % have to include label within this command
  \endgroup%
}
\makeatother

\newcommand\MyHighlight[1]{#1} % remove highlighting by uncommenting this line

\begin{document}

\bibliographystyle{apsrev4-1}% define bibliography style

\title{Cooperative eigenmodes and scattering in 1D atomic arrays}
\author{Robert J. Bettles}\email{r.j.bettles@durham.ac.uk}
\author{Simon A. Gardiner}\email{s.a.gardiner@durham.ac.uk}
\author{Charles S. Adams}\email{c.s.adams@durham.ac.uk}
\affiliation{Joint Quantum Center (JQC) Durham--Newcastle, Department of Physics,  Durham University, South Road, Durham, DH1 3LE, United Kingdom}
\date{\today}

\begin{abstract}
\MyHighlight{Collective coupling between dipoles can dramatically modify the optical response of a medium. Such effects depend strongly on the geometry of the medium and the polarization of the light. Using a classical coupled dipole model, here we investigate the simplest case of one dimensional (1D) arrays of interacting atomic dipoles driven by a weak laser field. 
}  
Changing the polarization and direction of the driving field allows us to separately address superradiant, subradiant, red-shifted, and blue-shifted eigenmodes, as well as observe strong Fano-like interferences between different modes. The cooperative eigenvectors can be characterized by the phase difference between nearest neighbor dipoles, ranging from all oscillating in phase to all oscillating out of phase with their nearest neighbors. Investigating the eigenvalue behavior as a function of atom number and lattice spacing, we find that certain eigenmodes of an infinite atomic chain have the same decay rate as a single atom between two mirrors.
\MyHighlight{The effects we observe provide a framework for collective control of the optical response of a medium, giving insight into the behavior of more complicated geometries, as well as providing further evidence for the dipolar analog of cavity QED. }
\end{abstract}


\maketitle

%%%%%%%%
\section{Introduction}

% cooperativity: effects and systems it has been observed in
The optical response of an ensemble of scatterers can be significantly modified if the scatterers behave \textit{cooperatively}, i.e.\ as an ensemble, rather than independently. Examples of cooperative effects can include enhanced and reduced scattering or decay rates (superradiance and subradiance respectively) \cite{Dicke1954,Stephen1964,Lehmberg1970,Friedberg1973,Gross1982,DeVoe1996,Inouye1999,Yoshikawa2005,Scheibner2007,Greenberg2012,Goban2015a,Guerin2016a,Araujo2016,Roof2016}, energy shifts \cite{Friedberg1973,Keaveney2012,Rohlsberger2010,Meir2014,Javanainen2014,Jennewein2015d,Jenkins2016,Roof2016}, highly directional scattering \cite{Rouabah2014,Oppel2014}, Fano-like interferences \cite{Luk'yanchuk2010,Ghenuche2012,Jenkins2013,Hopkins2013,Puthumpally-Joseph2014a,Bettles2015d} and modified optical depth and scattering \cite{Bettles2016,Chomaz2012,Pellegrino2014a,Kemp,Jennewein2015d}. Cooperative effects have been observed experimentally in many different systems, from ultracold (Bose-Einstein condensate) \cite{Inouye1999}, cold \cite{Yoshikawa2005,Greenberg2012,Goban2015a,Guerin2016a,Araujo2016,Jennewein2015d,Jenkins2016,Bromley2016,Roof2016} and high temperature atomic vapors \cite{Keaveney2012} to ions \cite{DeVoe1996,Meir2014,Casabone2015}, nuclei \cite{Rohlsberger2010}, quantum dots \cite{Scheibner2007} and plasmonic nanoresonators \cite{Luk'yanchuk2010,Ghenuche2012,Adamo2012,Ross2016a}. Understanding and being able to tailor this behavior may open the door to exciting and novel applications, including enhanced atom--light coupling \cite{Bettles2016}, shift-free clock transitions \cite{Kramer}, and long-lived quantum state storage \cite{Olmos2013}.  

% lattices
One way in which an ensemble can exhibit cooperative behavior is if the particles all scatter coherently \footnote{Cooperativity can also occur from incoherent sources, provided the scatterers are indistinguishable to the detectors \cite{Oppel2014}}. Such coherent scattering is a result partly of coherent driving by an external light field as well as coherent interactions between the particles (typically, electric dipole--dipole interactions). The combination of the resonant nature of these interactions as well as Bragglike interference between the scattered fields means that placing the scatterers into periodic arrays or lattices can greatly enhance the cooperative response \cite{Adamo2012,Meir2014,Jenkins2012,Jenkins2013,Nienhuis1987,Chang2012,Olmos2013,Bettles2015d,Bettles2016,Kramer,Kramer2015,Sutherland2016,Yoo2016a,Ross2016a}. This also has relevance for the study of spin lattices, since coherent scattering between two level dipoles maps exactly onto a spin exchange description \cite{Yan2013,Barredo2015,Olmos2013}. 
In two recent papers \cite{Bettles2015d,Bettles2016}, we investigated numerically the cooperative behavior of different two dimensional (2D) atomic lattices. \MyHighlight{In \cite{Bettles2015d} and also in \cite{Jenkins2013,Puthumpally-Joseph2014a}, strong Fano-like interferences between different cooperative eigenmodes can lead to a cooperative analog of electromagnetically induced transparency.} In \cite{Bettles2016}, we found certain parameter regimes in which the optical extinction through a 2D lattice can reach almost $100\%$, due in part to strong subradiant behavior of the dominant cooperative eigenmode (see also \cite{Jenkins2013,Ghenuche2012}). In this paper, we will discuss in more detail the model that was used in these previous works, and then apply it to the case of atoms trapped in one dimensional (1D) arrays.  
Investigation into the cooperative behavior of 1D arrays has already seen considerable interest in a number of different systems. One of the earliest measurements of the cooperative energy shifts and modified decay rates as a function of atom spacing was made for a pair of ions \cite{DeVoe1996}, which has more recently been extended to 1D arrays of up to 8 ions \cite{Meir2014}. Recent experiments have seen atoms coupled to 1D waveguides, in which superradiance has been observed \cite{Goban2015a}, localized eigenmodes and strong coupling predicted \cite{Haakh2015}, and optical band gaps and near perfect reflection predicted \cite{Chang2012,Liao2015} and recently measured \cite{Corzo2016,Sorensen2016a}. Other predictions for atoms coupling through free space include large energy shifts and modified decay rates as $N\to\infty$ \cite{Nienhuis1987,Kramer}, increase in excitation population along the direction of light propagation breaking the Beer-Lambert prediction \cite{Sutherland2016}, and subradiant excitation hopping \cite{Olmos2013} and state preparation \cite{Jen2016}.
%, and Bragg scattering and photonic band gaps by replacing each atom with a 2D disk of randomly positioned atoms \cite{Samoylova2014}. 
Reducing the dimensionality to 1D simplifies the behavior compared with the 2D arrays considered in \cite{Bettles2015d,Bettles2016}, making it easier to observe patterns and structures which in turn provide insight into the more complicated 2D behavior. Even in 1D however we still observe a rich variety of different cooperative phenomena.

% cavity analogue
The cooperative modification of an ensemble's optical response is analogous to the modified behavior of a single quantum emitter inside a cavity \cite{Eschner2001}. In both cases, the optical emission environment (i.e.\ the electromagnetic (EM) field mode structure) of a single emitter is modified by the presence of either a nearby mirror (in the cavity case) or an additional emitter (in the cooperative case). Recent proposals have suggested the reproduction of cavitylike effects in cooperative ensembles (without the need for a cavity), including atomic mirrors \cite{Bettles2016,Chang2012}, access to the strong coupling regime \cite{Chang2012}, and cavity-free lazing \cite{Svidzinsky2013}. In this paper, we show that the decay rate of a single atom in a cavity is equivalent to that of an infinite chain of atoms. Furthermore, the many atom system contains additional degrees of freedom compared with the cavity case, allowing for richer, more varied behavior.

% paper overview
In Section \ref{sec:CoupledDipoleModel} we present the coupled dipole model used to calculate the optical response of the atomic array to a weak classical driving field. We show how this can be used to calculate the scattering cross section of the system and relate this to the behavior of the cooperative eigenmodes. In Sec.\ \ref{sec:AtomCavity1}, we begin by calculating the decay rate of a single atom within a cavity. By replacing the cavity mirrors with a long chain of atoms on either side, we observe the same decay rate as for the single atom--cavity system. To better understand the behavior of the atom chain, in Sec.\ \ref{sec:N3AtomChain} we consider a smaller chain of just $N=3$ atoms, demonstrating Fano interferences between the eigenmodes, energy shifts, and superradiant and subradiant behavior, all accessible by tailoring the polarization and direction of the incident driving field. Increasing the atom number to $N=25$, in Sec.\ \ref{sec:N25AtomChain} we look closely at the eigenvectors and eigenvalues of a longer chain of atoms, discovering patterns in both. In Sec.\ \ref{sec:VaryingN} we compare chains of different atom numbers, finding that certain eigenvalues converge as $N\to\infty$ which, as we found in Sec.\ \ref{sec:AtomCavity1}, is equivalent to replacing the infinite chain of atoms with a single atom between two mirrors. 
\MyHighlight{We then finally in Sec.\ \ref{sec:ComparisonBetween1Dand2D} make comparison between the behavior of these 1D arrays with the 2D arrays considered in \cite{Bettles2015d,Bettles2016}.} 
In Sec.\ \ref{sec:Conclusions} we conclude our findings and present a brief outlook for future work.

%%%%%%%%
\section{Coupled Dipole Model} \label{sec:CoupledDipoleModel}

%%%%
\subsection{Coupled classical dipoles} \label{sec:CoupledClassicalDipoles}
We begin by considering an ensemble of $N$ atoms with two manifolds of energy states characterized by the angular momentum quantum number $J$. We assume there is a single ground state $\ket{J_g=0}\equiv\ket{g}$, separated by an energy of $E_{ge}=\hbar\omega_0$ from three degenerate excited states $\ket{J_e=1,m_{J_e}=\{0,\pm1\}}$, where $m_J$ is the projection quantum number of $J$, $\omega_0$ is the atomic transition frequency, and $\hbar$ the reduced Planck constant. Such a system could, for example, be realized in the triplet transitions of Sr \cite{Olmos2013,Zhou2010} or Yb \cite{Barber2008,Fukuhara2009a}.
Arrays of singly occupied atomic lattices can be created, for example, in optical lattice Mott insulators (e.g.\ demonstrated in Sr \cite{Stellmer2012} and Yb \cite{Fukuhara2009a}) or in arrays of dipole traps \cite{Nogrette2014,Lester2015a}.

The ground and excited states can be coupled by applying a \textit{driving} EM field to the atoms. For a sufficiently weak monochromatic laser beam, the electric field component can be described as a classical electric field $\boldsymbol{\mathcal{E}}_k$ oscillating with frequency $\omega=ck$ ($k=2\pi/\lambda$ is the wavenumber, $\lambda$ is the wavelength \footnote{We assume $\omega\gg\gamma_0$ and so for all detunings in this paper, ${\lambda\simeq\lambda_0=2\pi c/\omega_0}$.}, and $c$ is the speed of light) \cite{GardinerZoller2}, 
\begin{equation} \label{eq:EFieldField}
  \boldsymbol{\mathcal{E}}_k(\mathbf{r},t) = \mathop{\mathbf{E}_k(\mathbf{r})}\text{e}^{-\text{i}\omega t} + \mathop{\mathbf{E}_k^*(\mathbf{r})}\text{e}^{\text{i}\omega t},
\end{equation}
where $\mathbf{E}_k$ is the time-independent (complex) field in the rotating frame. We shall assume the driving field is a uniform field propagating with wavevector $\mathbf{k}$, amplitude $E_k$ and polarization $\polvec_k$, $\mathbf{E}_k(\mathbf{r}) = E_k\mathop{\text{e}^{\text{i}\mathbf{k}\cdot\mathbf{r}}}\polvec_k$. 
Which excited state is addressed by the driving field depends on the polarization of the driving field. For convenience, we will transform the excited states from the angular momentum projection basis into the Cartesian basis: 
\begin{align}
  \ket{x}&\equiv \frac{1}{\sqrt{2}}\big(\ket{J_e=1,m_{J_e}=1}+\ket{J_e=1,m_{J_e}=-1}\big),
\nonumber \\
  \ket{y}&\equiv \frac{\text{i}}{\sqrt{2}}\big(\ket{J_e=1,m_{J_e}=-1}-\ket{J_e=1,m_{J_e}=1}\big),
\nonumber \\
  \ket{z}&\equiv \ket{J_e=1,m_{J_e}=0}.
\end{align}
Each state can now be excited by a driving field with the corresponding linear polarization (e.g.\ $\polvec_x$ will couple the ground and $\ket{x}$ states).

The quantum dynamics of the $i$th atom from an ensemble of $N$ atoms can be described by that atom's density matrix \MyHighlight{$\rho_i\equiv\ket{\Psi_i}\bra{\Psi_i}$ (for single atom wavefunction $\ket{\Psi_i}$)}. We have taken the trace over the EM field parts of the quantum system and are assuming the many atom quantum state is a product state of the single atom states, $\rho_{\text{at}}=\bigotimes_i\rho_i$, where $\bigotimes_i$ is the tensor product over all atoms $i\in\{1\dots N\}$. If we assume the driving field amplitude $E_k$ is sufficiently weak such that we can ignore the excited state populations, $\rho_{i}^{\nu\nu}\simeq 0$ for $\nu\in\{x,y,z\}$, we then need only consider the behavior of the individual atomic coherences, $\rho_{i}^{g\nu}$. 

In this weak driving limit, the resulting many body optical Bloch equations describing the dynamics of the atomic coherences are equivalent to describing the atoms as classical, coupled, driven electric dipoles \cite{Javanainen1999,Svidzinsky2010,BettlesThesis}. The expectation of the (vector) electric dipole operator acting on atom $i$, $\boldsymbol{\mathcal{D}}_i$, is 
\begin{align}
  \langle \boldsymbol{\mathcal{D}}_i\rangle &= \mathop{\mathrm{Tr}}\left\{\rho_{\text{at}}\left(\sum_{\nu\in\{x,y,z\}}\mathbf{d}_{\nu g}\ket{\nu_i}\bra{g_i} + \mathbf{d}_{g\nu}\ket{g_i}\bra{\nu_i}\right)\right\}
  \nonumber \\
  &= \mathbf{d}_i\,\text{e}^{-\text{i}\omega t} + \text{c.c.},
\end{align}
where Tr is the trace over all atoms, ${\mathbf{d}_{\nu g}=\polvec_{\nu}\bra{\nu}\mathcal{D}\ket{g}}=\mathbf{d}_{g\nu}^*$ is the dipole matrix element in direction $\nu$, \MyHighlight{$\mathcal{D}$ is the scalar electric dipole operator}, $\text{c.c.}$ is the complex conjugate, and 
\begin{equation}
  \mathbf{d}_i\equiv \mathbf{d}_{g\nu}\,\rho_{i}^{\nu g}\,\text{e}^{\text{i}\omega t},
\end{equation}
is the electric dipole moment of atom $i$ in the same rotating frame as $\mathbf{E}_k$ in Eq.\ (\ref{eq:EFieldField}). In the steady state, an oscillating electric field $\mathbf{E}(\mathbf{r})$ results in an oscillating dipole moment in atom $i$, 
\begin{equation} \label{eq:dsingle}
  \mathbf{d}_i = \alpha\mathbf{E}(\mathbf{r}_i),
\end{equation} 
where $\mathbf{r}_i$ is the position of the atom, $\alpha=-\alpha_0\gamma_0/(\Delta+\text{i}\gamma_0)$ is the polarizability of a single two level atom, $\alpha_0=6\pi\varepsilon_0/k_0^3$, $k_0=\omega_0/c$ is the wavenumber for the resonant atomic transition, $\varepsilon_0$ is the permittivity of free space, $\gamma_0$ is half the natural atomic decay rate and $\Delta=\omega-\omega_0$ is the detuning of the driving field from the resonant atomic transition. Each oscillating electric dipole in turn radiates an oscillating electric field,
\begin{align} \label{eq:dipoleField}
  \mathbf{E}_i(\mathbf{r}) =& \mathop{\mathsf{G}(\mathbf{R}_i)}\mathbf{d}_i
\nonumber \\
  =& \mathop{\frac{3}{2\alpha_0}}\text{e}^{\text{i}kR_i}\bigg\{\bigg[\frac{1}{kR_i}
  +\frac{\text{i}}{(kR_i)^2}-\frac{1}{(kR_i)^3}\bigg]\,\mathbf{d}_i
\nonumber \\
  & -\bigg[\frac{1}{kR_i} +\frac{3\text{i}}{(kR_i)^2}-\frac{3}{(kR_i)^3}\bigg]
  \left(\hat{\mathbf{R}}_i\cdot\mathbf{d}_i\right)\hat{\mathbf{R}}_i \bigg\},
\end{align}
where $\mathsf{G}(\mathbf{r})$ is a $3\times 3$ matrix with matrix elements ${(\nu,\upsilon)\in\{x,y,z\}}$, defined as above, and the vector ${\mathbf{R}_i =R_i\hat{\mathbf{R}}_i\equiv \mathbf{r}-\mathbf{r}_i}$ has magnitude $R_i$ and unit vector $\hat{\mathbf{R}}_i$ (we just consider $\mathbf{R}_i\neq0$).
$\mathsf{G}(\mathbf{r})$ is the Green's function solution for an electric dipole radiating into free space \cite{Jackson1963}. 
The total field experienced by atom $i$ is therefore the sum of the driving field and the fields scattered from every other dipole,
\begin{equation}
  \mathbf{E}(\mathbf{r}_i) = \mathbf{E}_k(\mathbf{r}_i) + \sum_{j\neq i}\mathop{\mathsf{G}_{ij}}\mathbf{d}_j,
\end{equation}
where $\mathsf{G}_{ij}\equiv\mathsf{G}(\mathbf{r}_i-\mathbf{r}_j)$ is the $3\times 3$ Green's function matrix describing the scattering between atoms $i$ and $j$. Substituting this into the expression for the dipole moment (\ref{eq:dsingle}), we obtain a set of $3N$ coupled linear equations:
\begin{equation} \label{eq:coupledDipoleEq}
  \mathbf{d}_i = \alpha\mathbf{E}_k(\mathbf{r}_i) + \alpha\sum_{j\neq i}\mathop{\mathsf{G}_{ij}}\mathbf{d}_j,
\end{equation}
where each vector $\mathbf{d}_i$ has three components.
These coupled equations can be numerically solved self-consistently, allowing us to calculate the steady state behavior of the dipole moments of an ensemble of atoms with arbitrary positions driven by a classical driving field with arbitrary polarization and functional form (in deriving (\ref{eq:coupledDipoleEq}) we have made no assumption on the atomic position or form of the electric field). Again, we emphasize that this is both the solution to the many body optical Bloch equations in the weak driving limit and equivalently the solution if we had modeled each atom as a classical coupled driven oscillator \cite{Javanainen1999,Svidzinsky2010,BettlesThesis}. This type of coupled dipole model has been used extensively in several fields including nanoplasmonics \cite{Hopkins2013,Jenkins2012a,Jenkins2013,Ross2016a} and atomic physics \cite{Morice1995,Ruostekoski1997a,Chomaz2012,Bettles2015d,Bettles2016,Araujo2016,Roof2016,Jennewein2015d,Jenkins2016,Bromley2016,Yoo2016a,Samoylova2014}. 

\MyHighlight{
By solving Eq.\ (\ref{eq:coupledDipoleEq}) self consistently, we are accounting for multiple recurrent scattering between the dipoles. The resulting phenomenology is different to if we had instead treated the atoms as a polarizable medium experiencing mean local field corrections, which is the case when, e.g., there is significant inhomogeneous broadening \cite{Javanainen2014}. In such systems, mean-field density-dependent phenomena can include collisional self-broadening of absorption lines \cite{Weller2011} and collective Lamb shifts \cite{Friedberg1973,Keaveney2012,Rohlsberger2010}. 
}

%%%%
\subsection{Eigenvalue decomposition}
The coupled linear equations in (\ref{eq:coupledDipoleEq}) can be represented in terms of a matrix equation $\vec{\mathbf{E}}_k=\mathsf{M}\,\vec{\mathbf{d}}$, where $\vec{\mathbf{E}}_k$ and $\vec{\mathbf{d}}$ are column vectors composed of the $N$ driving field and dipole vectors respectively, and $\mathsf{M}$ is a $3N\times 3N$ matrix describing the coupling between these vectors. $\mathsf{M}$ is composed of smaller $3\times 3$ submatrices, $\mathsf{M}_{ij}$, each describing the coupling between atoms $i$ and $j$. Each element of $\mathsf{M}_{ij}$ in turn describes the coupling between polarizations $(\nu,\upsilon)\in\{x,y,z\}$. The matrix elements of $\mathsf{M}$ therefore have the form
\begin{equation}
  \mathsf{M}_{ij}^{\nu\upsilon} = \alpha^{-1}\delta_{\nu,\upsilon}\,\delta_{i,j} - (1-\delta_{i,j})\,\mathsf{G}_{ij}^{\nu\upsilon},
\end{equation}
where $\mathsf{G}_{ij}^{\nu\upsilon}$ is the $(\nu,\upsilon)^{\text{th}}$ element of $\mathsf{G}_{ij}$. 

It is instructive to consider the eigenvalues $\mu_{\ell}$ and eigenvectors $\vec{\mathbf{m}}_{\ell}$ of the matrix $\mathsf{M}$, as this will provide insight into the behavior of $\vec{\mathbf{d}}$ \cite{Hopkins2013,Bettles2015d}. The eigenvalue equation for $\mathsf{M}$ is $\mathsf{M}\mathop{\vec{\mathbf{m}}_{\ell}} = \mu_{\ell}\mathop{\vec{\mathbf{m}}_{\ell}}$, where the eigenmode index is $\ell\in\{1\dots 3N\}$. Provided $\mathsf{M}$ is invertible, the set of eigenvectors $\{\vec{\mathbf{m}}_{\ell}\}$ forms a complete basis \footnote{A discussion of the conditions for when $\mathsf{M}$ is and is not invertible are left to later work.}. The tensors $\vec{\mathbf{E}}_k$ and $\vec{\mathbf{d}}$ can therefore be represented in terms of this eigenbasis:
\begin{subequations}\label{eq:EvecdvecEmodeExpansion} \begin{align} 
  \vec{\mathbf{E}}_k &= \sum_{\ell}b_{\ell} \mathop{\vec{\mathbf{m}}_{\ell}}, \label{eq:EFieldExpansion} \\
  \vec{\mathbf{d}} &= \sum_{\ell}c_{\ell} \mathop{\vec{\mathbf{m}}_{\ell}},  
\end{align} \end{subequations}
where the coefficients can be calculated by taking the dot product of (\ref{eq:EvecdvecEmodeExpansion}) with $\vec{\mathbf{m}}_{\ell'}$. If $\mathsf{M}$ were Hermitian, the eigevectors would be orthogonal and calculating the coefficients would be trivial. However, because the dipole--dipole coupling is complex and symmetric under exchange of atom \MyHighlight{and/or} polarization index, ${\mathsf{G}_{ij}^{\nu\upsilon}=\mathsf{G}_{ji}^{\upsilon\nu}\neq(\mathsf{G}_{ji}^{\upsilon\nu})^*}$, the matrix $\mathsf{M}$ is \textit{complex symmetric}, rather than Hermitian. One consequence of this is that the eigenvectors are not necessarily orthogonal, i.e.\ there are situations when $\vec{\mathbf{m}}_{\ell}^{*}\cdot\vec{\mathbf{m}}_{\ell'}\neq\delta_{\ell,\ell'}$ \footnote{We define our notation for the dot product of complex column vectors as $\vec{\mathbf{a}}^*\cdot\vec{\mathbf{b}}\equiv\sum_p a_p^* b_p$, where $^*$ denotes a complex conjugate.}. Calculating each coefficient $b_{\ell}$ then involves solving a set of coupled linear equations:
\begin{align}
  \vec{\mathbf{m}}_{\ell}^{*}\cdot\vec{\mathbf{E}}_k = b_{\ell} + \sum_{\ell'\neq\ell} b_{\ell'}\,\vec{\mathbf{m}}_{\ell}^{*}\cdot\vec{\mathbf{m}}_{\ell'},
\end{align}
assuming we have normalized $|\vec{\mathbf{m}}_{\ell}|^2 = 1$. From this, we can calculate the expansion coefficients for $\vec{\mathbf{d}}$, 
\begin{align}
  \vec{\mathbf{E}}_k =& \mathsf{M}\mathop{\vec{\mathbf{d}}} 
  = \mathsf{M}\sum_{\ell}c_{\ell}\mathop{\vec{\mathbf{m}}_{\ell}}
  = \sum_{\ell} c_{\ell} \,\mu_{\ell}\mathop{\vec{\mathbf{m}}_{\ell}} 
  = \sum_{\ell} b_{\ell} \mathop{\vec{\mathbf{m}}_{\ell}},
\end{align}
i.e., $c_{\ell}=b_{\ell}/\mu_{\ell}$ and $\vec{\mathbf{d}} = \sum_{\ell}b_{\ell}\vec{\mathbf{m}}_{\ell}/\mu_{\ell}$.

One further consequence of the complex symmetry of $\mathsf{M}$ is that the eigenvalues are, in general, complex. The interaction energy between two electric dipoles $\mathbf{d}_i$ and $\mathbf{d}_j$ is given by 
\begin{equation} \label{eq:Vdd}
  V_{\text{dd}}=-\mathbf{d}_j^*\cdot\mathsf{G}_{ji}\,\mathbf{d}_i=-\mathbf{d}_i^*\cdot\mathsf{G}_{ij}\,\mathbf{d}_j.
\end{equation}
The complex nature of $\mathsf{G}$ is related to how it has both a coherent and dissipative part. If we split the coupling matrix into the diagonal matrix $\mathbb{1}/\alpha$ and the coupling matrix $\mathsf{G}$ (where $\mathsf{G}$ is the $3N\times 3N$ matrix with matrix elements $\mathsf{G}_{ij}^{\nu\upsilon}$), i.e.\ $\mathsf{M}=\mathbb{1}/\alpha-\mathsf{G}$, then the eigenvalues can be expressed as
\begin{align} \label{eq:evalComplexExpansion}
  \mu_{\ell} =& \frac{1}{\alpha} - g_{\ell}
  = -\frac{1}{\alpha_0}\frac{\Delta+\text{i}\gamma_0}{\gamma_0} - g_{\ell}
\nonumber \\
  =& -\frac{1}{\alpha_0\gamma_0}\left[(\Delta-\Delta_{\ell})+\text{i}(\gamma_0+\gamma_{\ell})\right],
\end{align}
where $g_{\ell}$ is the eigenvalue of the coupling matrix $\mathsf{G}$, and $\Delta_{\ell}$ and $\gamma_{\ell}$ are related to the real and imaginary parts of $g_{\ell}$ respectively, 
\begin{align}
  \Delta_{\ell}\equiv-\alpha_0\mathop{\gamma_0}\mathop{\mathrm{Re}}(g_{\ell}), &&
  \gamma_{\ell}\equiv\alpha_0\mathop{\gamma_0}\mathop{\mathrm{Im}}(g_{\ell}).
\end{align}
The eigenvalues in Eq.\ (\ref{eq:evalComplexExpansion}) have a similar form to the inverse of the atomic polarizability $\alpha^{-1}=-(\Delta+\text{i}\gamma_0)/(\alpha_0\gamma_0)$, except the resonance frequency is shifted by $\Delta_{\ell}$ and the decay rate is modified by $\gamma_{\ell}$. Because of this, we shall refer to $\Delta_{\ell}$ as the \textit{cooperative shift} and $(\gamma_{0}+\gamma_{\ell})$ as the \textit{cooperative (half) decay rate}. When $(\gamma_0+\gamma_{\ell})>\gamma_0$, the decay rate is said to be \textit{superradiant}; when $(\gamma_0+\gamma_{\ell})<\gamma_0$, the decay rate is said to be \textit{subradiant}.

%%%%
\subsection{Degenerate eigenmodes} 
In this paper we consider atoms arranged in 1D arrays. If we define three orthogonal coordinate axes such that one is parallel to the atomic separation vector (the \textit{atomic axis}), $\polvec_{\parallel}$, and the other two are perpendicular to the atomic axis, $\polvec_{\perp,1}$ and $\polvec_{\perp,2}$, then Eq.\ (\ref{eq:dipoleField}) shows that there is no dipole--dipole interaction between dipoles that are aligned along different axes from $\{\polvec_{\parallel},\polvec_{\perp,1},\polvec_{\perp,2}\}$. We can therefore separate the $3N$ eigenmodes equally into modes with dipoles polarized separately in each of these three coordinate axes.

\MyHighlight{For a given polarization $\polvec$, the angle between $\hat{\mathbf{R}}_{ij}$ and $\polvec$ is $\theta=\arccos(\hat{\mathbf{R}}_{ij}\cdot\polvec)$.}
Because $\theta=\pi/2$ is the same for both $\polvec_{\perp,1}$ and $\polvec_{\perp,2}$, each mode in $\polvec_{\perp,1}$ is degenerate with an identical mode in $\polvec_{\perp,2}$. For two degenerate eigenvectors $\vec{\mathbf{m}}_{\ell}$ and $\vec{\mathbf{m}}_{\ell'}$, any linear superposition of these two eigenvectors is also an eigenvector of $\mathsf{M}$ with the same eigenvalue. To speed up calculations, if we consider a driving field that only excites one of the three mentioned polarizations, then we can ignore the other two directions in our calculations.

%%%%
\subsection{Scattering cross section} \label{sec:scatteringCS}
One convenient quantity we can calculate from the dipole solutions is the \textit{scattering cross section}. The scattering cross section for an ensemble of electric dipoles is given by
\begin{equation} \label{eq:scatteringCS}
  \sigma_{\text{sc}} = \frac{\sigma_0}{\alpha_0|E_k|^2} \mathop{\mathrm{Im}}\left(
  \vec{\mathbf{E}}_k^*\cdot\vec{\mathbf{d}}\right),
\end{equation}
where $\sigma_0=6\pi/k_0^2$ is the resonant atomic scattering cross section. We assume the atomic dipoles have no nonradiative decay (e.g.\ phonon loss in a plasmonic resonator), and so the scattering cross section is equal to the extinction cross section, which can be determined using the optical theorem \cite{Jackson1963}. Substituting the expressions for $\vec{\mathbf{E}}_k$ and $\vec{\mathbf{d}}$ from (\ref{eq:EvecdvecEmodeExpansion}) into (\ref{eq:scatteringCS}) gives
\begin{align} \label{eq:scatteringCSDecomposition}
  \sigma_{\text{sc}} =& \frac{\sigma_0}{\alpha_0|E_k|^2}   
  \mathop{\mathrm{Im}}\left[\left( 
  \sum_{\ell}b_{\ell}^*\mathop{\vec{\mathbf{m}}_{\ell}^*}
  \right)\cdot\left(
  \sum_{\ell'}\frac{b_{\ell'}}{\mu_{\ell'}}\mathop{\vec{\mathbf{m}}_{\ell'}}
  \right)\right]
\nonumber \\
  =&  \frac{\sigma_0}{\alpha_0|E_k|^2} \left[
  \sum_{\ell}|b_{\ell}|^2 \mathop{\mathrm{Im}}\left(\frac{1}{\mu_{\ell}}\right) +
  \sum_{\ell,\ell'}^{\ell\neq\ell'} \mathop{\mathrm{Im}}\left( 
  \frac{b_{\ell}^*b_{\ell'}}{\mu_{\ell'}} \vec{\mathbf{m}}_{\ell}^*\cdot
  \vec{\mathbf{m}}_{\ell'} \right) \right].
\end{align}
For clarity we define the terms in the sum just over $\ell$ as \textit{direct} contributions to the cross section, $\sigma_{\ell}$, and the terms in the sum over $\ell$ and $\ell'$ as \textit{interference} contributions, $\sigma_{\ell\ell'}$, i.e.\
\begin{equation}
  \sigma_{\text{sc}} \equiv \sum_{\ell}\sigma_{\ell} + \sum_{\ell,\ell'}^{\ell\neq\ell'} \sigma_{\ell\ell'}, 
\end{equation}
The significance of the nonorthogonality of the eigenvectors for the cross section is that not only does each mode contribute to the scattering individually ($\sigma_{\ell}$), but there are also interferences between modes ($\sigma_{\ell\ell'}$), which, as we shall see in Sec.\ \ref{sec:N3AtomChain}, result in striking Fano-like resonance interferences. The direct scattering due to each mode has a Lorentzian lineshape:
\begin{equation}
  \sigma_{\ell} = \frac{\sigma_0|b_{\ell}|^2}{|E_k|^2}\frac{\gamma_0(\gamma_0+\gamma_{\ell})}{(\Delta-\Delta_{\ell})^2 + (\gamma_0+\gamma_{\ell})^2},
\end{equation}
which has a resonance shifted by $\Delta_{\ell}$, a half-width--half-maximum (HWHM) of $(\gamma_0+\gamma_{\ell})$, and a value on resonance of 
\begin{equation}
  \sigma_{\ell}(\Delta=\Delta_{\ell}) = \frac{\sigma_0|b_{\ell}|^2}{|E_k|^2}
  \frac{\gamma_0}{\gamma_0+\gamma_{\ell}},
\end{equation}
which is inversely proportional to the ratio of the cooperative decay rate and the natural decay rate. A superradiant resonance ($\gamma_0+\gamma_{\ell}>\gamma_0$) will therefore broaden and lower the peak of the lineshape of $\sigma_{\ell}$, whilst a subradiant resonance will narrow and increase the peak of the lineshape.

%%%%%%%%
\section{Single atom in a cavity} \label{sec:AtomCavity1}

Before investigating in detail the cooperative behavior of different 1D atomic chains, we want to make a comparison between the way multiple atoms interact with each other to the way a single atom interacts with a mirror. 
To modify the optical response of a single resonator, it is necessary to modify the EM environment of that resonator. For a single atom, this can be done, for example, by placing the atom within an optical cavity (e.g.\ between two highly reflecting mirrors). The EM field generated in the mirror surface is equivalent to there being an image dipole positioned behind the mirror with which the real dipole can then interact \cite{Jackson1963}.
Placing the atom midway between two mirrors separated by $a$ therefore results in the real dipole interacting with an infinite chain of equally spaced image dipoles. If the dipole is polarized parallel to the mirror planes, the first order image dipoles on either side of the real dipole are antialigned with the real dipole, and the half-decay rate is \cite{Milonni1973}
\begin{equation} \label{eq:GammaPara}
  \gamma^{\parallel} = \frac{3\pi\gamma_0}{2ka}\sum_{n=1}^{ka/\pi}\left(
  1+\frac{n^2\pi^2}{k^2a^2}\right)\sin^2\left(\frac{n\pi}{2}\right),
\end{equation}
where $n$ is the cavity mode index.
Alternatively, if the real dipole is polarized perpendicular to the mirror planes, the image dipoles are then all aligned with the real dipole, and the half-decay rate is
\begin{equation} \label{eq:GammaPerp}
  \gamma^{\perp} = \frac{3\pi\gamma_0}{ka}\left[\frac{1}{2} +
  \sum_{n=1}^{ka/\pi}\left(1-\frac{n^2\pi^2}{k^2a^2}\right)
  \cos^2\left(\frac{n\pi}{2}\right) \right].
\end{equation}
The number of EM modes that can exist within the cavity is limited by the size of the cavity.
%, $n\leq ka/\pi$. 
For the parallel polarization $\gamma_{\parallel}$ plotted in Fig.\ \ref{fig:NAtomChainShiftsWidthsInfSingle}, if the cavity is too short to support even a single cavity mode ($a<\lambda/2$) then the atom cannot decay and so the decay rate becomes zero (subradiance). This is because the cavity mode must have opposite sign at the real and image dipoles. Conversely, for $\gamma_{\perp}$, we shall see later in Fig.\ \ref{fig:NAtomChainShiftsWidthsInf}(b) that the atom can decay even when $a<\lambda/2$ and, in fact, $\gamma_{\perp}\to\infty$ (superradiance) as $a\to 0$.

% atom in a cavity plot
\begin{figure}
\begin{center}
\includegraphics[width=8.6cm]{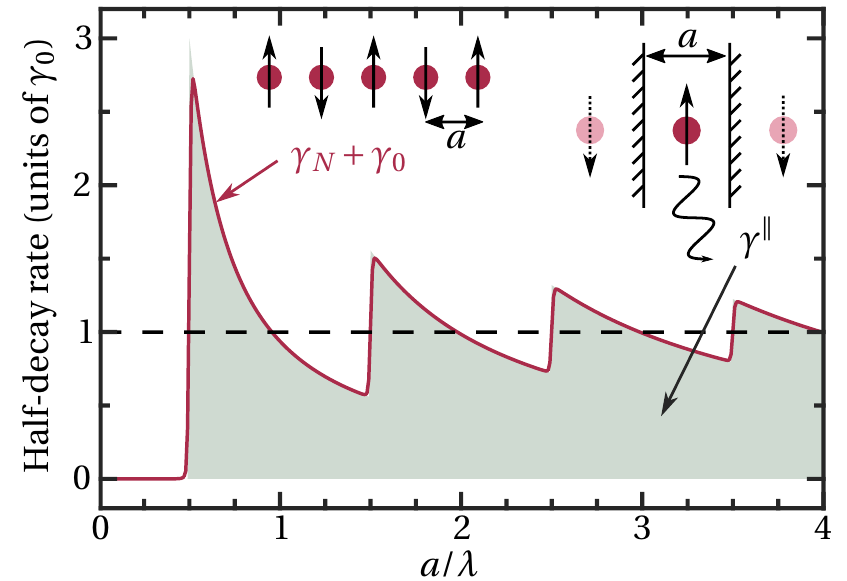}
\colorcaption{Half-decay rate for a single atom between two mirrors and polarized parallel to the mirror plane [grey shaded area, Eq.\ (\ref{eq:GammaPara})] as a function of mirror spacing $a$. This is shown to have a similar decay rate behavior to that of a chain of $N=51$ real atoms polarized perendicular to the atom chain, antialigned with their nearest neighbors (red solid line), and separated by nearest neighbor spacing $a$. 
}{fig:NAtomChainShiftsWidthsInfSingle}
\end{center}
\end{figure}

Using the model outlined in Sec.\ \ref{sec:CoupledDipoleModel}, we can replace the image dipoles formed by the mirrors with a chain of real dipoles. In Fig.\ \ref{fig:NAtomChainShiftsWidthsInfSingle} we see that the cavity half-decay rate is approximated well by the half-decay rate of a chain of $N=51$ atoms polarized perpendicular to the atomic axis and in the eigenmode for which each dipole is antialigned with its nearest neighbors ($\vec{\mathbf{m}}_{N}$, the mode index is $\ell=N$ which will be explained in Sec.\ \ref{sec:N25AtomChain}). The chain of atoms therefore behaves as if the atoms on either side of the central atom are just mirrors, allowing only certain modes to be supported. Similar mirrorlike behavior has been predicted \cite{Chang2012} and recently demonstrated \cite{Corzo2016,Sorensen2016a} for 1D chains of atoms coupled along a waveguide, where Bragg reflection from the atom chains can, in an ideal case, produce near perfect reflection of an incident electric field propagating through the waveguide. The dipolar system therefore provides an analog to cavity QED, although with additional degrees of freedom since the behavior of each dipole is no longer constrained by the behavior of the central dipole, as is the case with the image dipoles. In the following Sections we shall go on to investigate this cooperative behavior, looking at the scattering, eigenvectors and eigenvalues for different 1D atomic chains.

%%%%%%%%
\section{Atom chain, $N=3$} \label{sec:N3AtomChain}

%%%%
\subsection{Perpendicular wavevector, parallel polarization}
In order to better understand the behavior of the chain of dipoles shown in Fig.\ \ref{fig:NAtomChainShiftsWidthsInfSingle}, we shall now consider a much simpler system of just three atoms in a chain. Such a system has also been considered in \cite{Feng2013a}.
In Fig.\ \ref{fig:ThreeAtomsExtinctionModes} we plot the scattering cross section as a function of detuning for three different orientations of driving field polarization $\polvec_k$ and wavevector $\mathbf{k}$. 
In Fig.\ \ref{fig:ThreeAtomsExtinctionModes}(a) we first consider the case where the driving field is incident perpendicular to the chain, $\mathbf{k}_{\perp,2}$, and polarized parallel to the chain, $\polvec_{\parallel}$. The overall scattering cross section (red solid line) exhibits a broadening and a red shift of the resonance lineshape compared with the single atom case (grey shaded area). However, in addition to this, the lineshape has a very sharp blue-shifted resonance. 

The presence of these two features can be explained by considering the eigenmode decomposition of $\sigma_{\text{sc}}$ (\ref{eq:scatteringCSDecomposition}). By plotting the individual cross sections of the different modes, $\sigma_{\ell}$, we see that the overall lineshape is dominated by two individual modes, one broad (superradiant, $\gamma_{1'}\simeq2.25\gamma_0$) and red shifted ($\sigma_{1'}$, green line), and the other narrow (subradiant, $\gamma_{3'}\simeq0.06\gamma_0$) and blue shifted ($\sigma_{3'}$, blue line) \footnote{The mode numbering is chosen so as to be compatible with the mode numbering in Fig.\ \ref{fig:NAtomsChainModeVectors}. The prime indicates modes polarized parallel to the atomic axis.}. The eigenvector for mode $\sigma_{1'}$ corresponds to each dipole oscillating with approximately equal amplitude and approximately in phase with each other ($\mathbf{d}_{1,3}\simeq0.7\mathop{\mathrm{exp}}(0.02\text{i}\pi)\,\mathbf{d}_2$ for edge dipoles $\mathbf{d}_{1,3}$ and central dipole $\mathbf{d}_{2}$). In Sec.\ \ref{sec:N25Eigenvalues} we will discuss how, for small spacing, these modes are similar to the well known Dicke states \cite{Dicke1954}. 

The total cross section is not, however, just the sum of the two mode cross sections $\sigma_{1'}+\sigma_{3'}$. In Fig.\ \ref{fig:ThreeAtomsExtinctionModes}(b) we plot the difference in the sum of the two mode cross sections with the total cross section $(\sigma_{\mathrm{sc}}-\sigma_{1'}-\sigma_{3'})$, which is identical to the interference term $\sigma_{\ell\ell'}$. This interference, as already mentioned, is asymmetric around the resonance of mode $\vec{\mathbf{m}}_{3'}$. Such an asymmetric interference lineshape is characteristic of a Fano-type lineshape. In a Fano resonance, a discrete ground state can be excited to a continuum of excited states either directly or via an intermediate discrete state, the energy of which lies within the excited state energy band. Interference between these two pathways changes sign as the frequency of the driving passes through resonance with the discrete state, resulting in the characteristic Fano asymmetric lineshape. In the coupled atomic dipoles, excitation from the ground state to a broad cooperative eigenstate ($\vec{\mathbf{m}}_{1'}$) can either occur directly or via the narrow eigenstate ($\vec{\mathbf{m}}_{3'}$). This pathway is allowed because the eigenvectors are nonorthogonal. As with the Fano resonance, the sign of the interference changes as the driving goes through resonance with the narrow mode, resulting in an asymmetric interference lineshape. Fano-like interference lineshapes have been predicted and observed in a number of coupled dipole systems \cite{Hopkins2013,Hopkins2015,Bettles2015d,Luk'yanchuk2010,Chong2014,Emami2014}.

% scattering cross section and modes for N=3
\begin{figure}[t]
\begin{center}
\includegraphics[width=8.6cm]{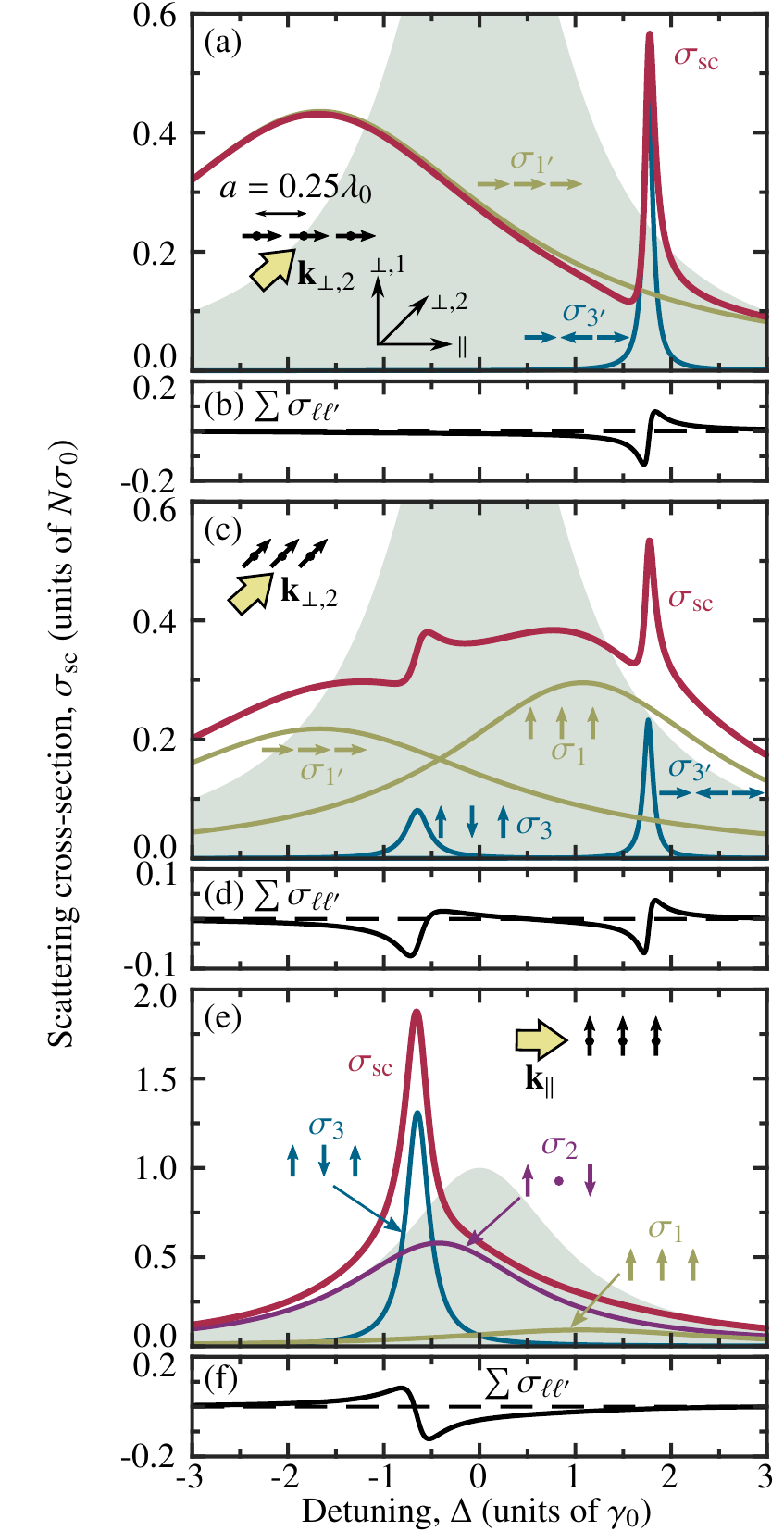}
\colorcaption
{(a,c,e) Scattering cross section $\sigma_{\text{sc}}$ (red solid lines) of a uniform, linearly polarized plane wave due to three atoms evenly spaced in a line with nearest neighbor spacing $a=0.25\lambda$, as a function of the driving field detuning. The contribution to the scattering from the individual modes $\sigma_{\ell}$ are also plotted (blue, green and purple solid lines). 
(b,d,f) The solid black lines plot the contribution to the scattering from interference between the modes, $\sum_{\ell\ell'}^{\ell\neq\ell'}\sigma_{\ell\ell'}$. 
(a,b) The driving field wavevector is perpendicular to the atomic axis, $\mathbf{k}_{\perp,2}$, and polarized parallel to the atomic axis, $\polvec_{\parallel}$. (c,d) The driving field wavevector is perpendicular to the atomic axis, $\mathbf{k}_{\perp,2}$, and linearly polarized $\pi/4$ to the atomic axis, $\polvec=(\polvec_{\parallel}+\polvec_{\perp,1})/\sqrt{2}$. (e,f) The driving field wavevector is parallel to the atomic axis, $\mathbf{k}_{\parallel}$, and polarized perpendicular to the atomic axis, $\polvec_{\perp,1}$.
The grey shaded areas indicate the scattering lineshape for a single noninteracting atom.
}{fig:ThreeAtomsExtinctionModes}
\end{center}
\end{figure} 

%%%%
\subsection{Perpendicular wavevector, diagonal polarization}
If we change the angle of the polarization vector such that it is diagonal with equal components in $\polvec_{\parallel}$ and $\polvec_{\perp,1}$, then we excite twice as many eigenstates [Fig.\ \ref{fig:ThreeAtomsExtinctionModes}(c)]. In addition to the two $\polvec_{\parallel}$ states observed in Fig.\ \ref{fig:ThreeAtomsExtinctionModes}(a), we observe the equivalent in phase ($\vec{\mathbf{m}}_1$) and out of phase ($\vec{\mathbf{m}}_3$) modes polarized in $\polvec_{\perp,1}$. The eigenvalues of the $\polvec_{\perp,1}$ modes are different to those of the $\polvec_{\parallel}$ modes because the dipole--dipole interaction energy is different for $\theta=0$ and $\theta=\pi/2$. $\sigma_1$ is again broad (superradiant) and $\sigma_{3}$ is narrow (subradiant), although the shifts now have opposite sign. Like $\vec{\mathbf{m}}_{1'}$ and $\vec{\mathbf{m}}_{3'}$, $\vec{\mathbf{m}}_{1}$ and $\vec{\mathbf{m}}_{3}$ interfere with each other, resulting in an asymmetric interference lineshape in Fig.\ \ref{fig:ThreeAtomsExtinctionModes}(d) at the resonance of $\sigma_{3}$. The two sets of modes of different polarizations do not however interfere as they truly are orthogonal (i.e.\ $\vec{\mathbf{m}}_{1}^*\cdot\vec{\mathbf{m}}_{1'}=0$, etc.), and so the only interferences occur between modes with nonorthogonal polarization. 

% eigenvectors for N=25
\begin{figure*}[t]
\begin{center}
\includegraphics[width=12.7cm]{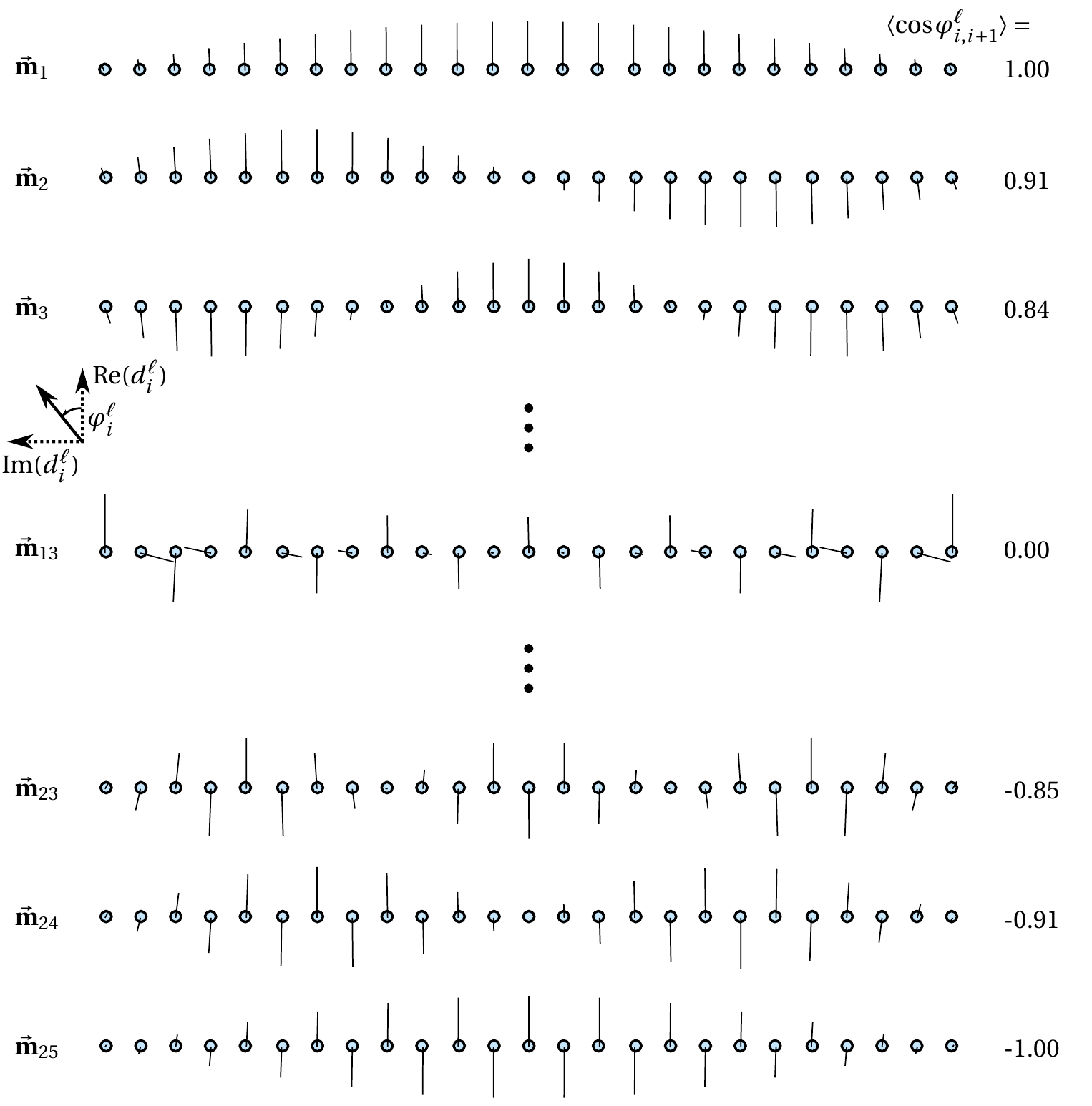}
\colorcaption
{Phasors of each dipole in a chain of $N=25$ atoms for a selection of eigenmodes. 
Each phasor represents the amplitude $d_i^{\ell}$ and phase $\varphi_i^{\ell}$ of each dipole in the $\ell$th eigenmode polarized perpendicular to the atomic axis ($\theta=\pi/2$), $\mathbf{d}_i^{\ell}=d_i^{\ell}\mathop{\text{e}}^{\text{i}\varphi_i^{\ell}}\polvec_{\perp,1}$. Atomic spacing is $a=0.25\lambda$. The nearest neighbour phase correlation function $\langle\cos\varphi_{i,i+1}^{\ell}\rangle$, Eq.\ (\ref{eq:phiPhaseCorrelationFunc}), decreases with increasing mode index. 
}{fig:NAtomsChainModeVectors}
\end{center}
\end{figure*}

%%%%
\subsection{Parallel wavevector, perpendicular polarization}
Changing the direction of propagation of the driving field in Fig.\ \ref{fig:ThreeAtomsExtinctionModes}(e) such that the propagation wavevector is parallel to the atomic axis, we are able to excite a third mode not previously seen in either of the other configurations: $\vec{\mathbf{m}}_{2}$. This mode corresponds to the central atom having no dipole moment whilst the outer two dipoles oscillate with equal amplitude and $\pi$ out of phase with each other. The reason this \textit{antisymmetric} mode is not observed in the other two cases is due to the symmetry of the driving field: in Fig.\ \ref{fig:ThreeAtomsExtinctionModes}(a--d), the driving field experienced by each atom is identical, and therefore the overlap between mode $\sigma_2$ and these fields is zero, meaning $b_2=0$. For $\mathbf{k}_{\parallel}$, however, the propagation phase $\text{e}^{\text{i}\mathbf{k}_{\parallel}\cdot \mathbf{r}}$ can be different at each atom, meaning that depending on the value of $a$, each atom experiences a different phased driving field. This means there can now be a nonzero overlap with an antisymmetric mode like $\vec{\mathbf{m}}_2$. For $a=0.25\lambda$, the phase difference between each nearest neighbor is $\text{e}^{\text{i}\pi/2}=\text{i}$, which results in the expansion coefficient of mode $\vec{\mathbf{m}}_2$, $|b_2|^2$ (\ref{eq:EFieldExpansion}), being around four times larger than $|b_1|^2$ and $|b_3|^2$. Because the linewidth of $\sigma_3$ is so narrow, however, the peak of $\sigma_3$ is still larger than the peak of $\sigma_2$. Note, as well, that because mode $\vec{\mathbf{m}}_{2}$ is orthogonal to $\vec{\mathbf{m}}_{1}$ and $\vec{\mathbf{m}}_{3}$, the only nonzero interferences in Fig.\ \ref{fig:ThreeAtomsExtinctionModes}(f) are between modes $\vec{\mathbf{m}}_{1}$ and $\vec{\mathbf{m}}_{3}$. 

%%%%
\subsection{Comment on application}
Even for just three atoms with fixed atomic spacing, we observe a diverse range of different scattering behaviors. Depending on what is required, we can realize different features just by changing the direction and polarization of the driving field. For example, Fig.\ \ref{fig:ThreeAtomsExtinctionModes}(a) allows us to observe strong mode interferences and simultaneous superradiant and subradiant behavior, depending on detuning. In Fig.\ \ref{fig:ThreeAtomsExtinctionModes}(c), the lineshape is dominated by two orthogonal broad superradiant modes which do not interfere and so the overall lineshape is broad and superradiant, with only relatively weak contributions from the two subradiant modes. Conversely, in Fig.\ \ref{fig:ThreeAtomsExtinctionModes}(e), simultaneous excitation of symmetric and antisymmetric modes results in only weak mode interferences and strong excitation of a subradiant mode, meaning the overall lineshape is now largely subradiant, with the peak cross section almost doubling that of the independent atom case.

%%%%%%%%
\section{Atom chain, $N=25$} \label{sec:N25AtomChain}
%%%%
\subsection{Eigenvectors}

% correlation functions of the eigenvectors in N=25
\begin{figure}[t]
\begin{center}
\includegraphics[width=8.6cm]{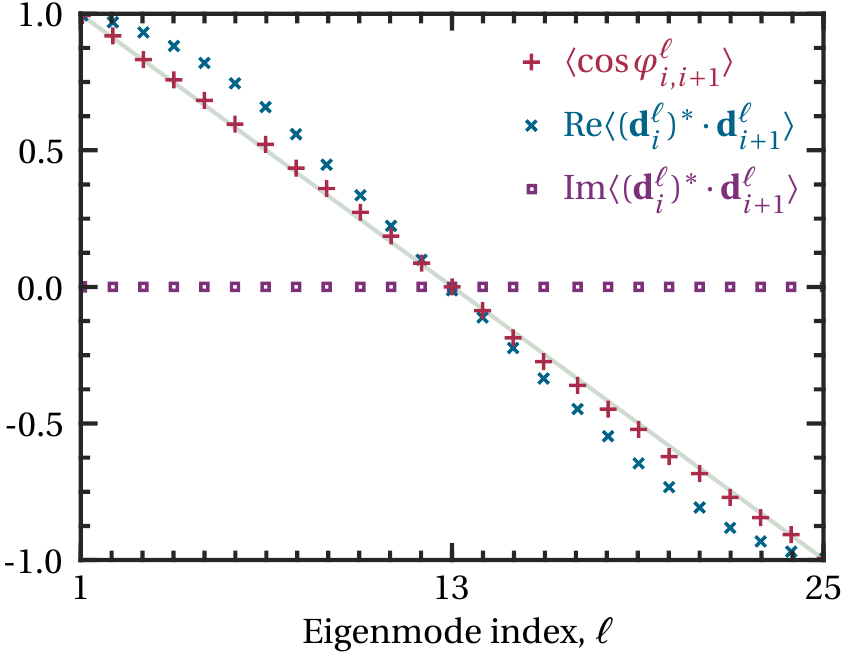}
\colorcaption
{Nearest neighbor phase correlation functions for each eigenmode in a chain of $N=25$ atoms polarized perpendicular to the atomic axis ($\theta=\pi/2$) with nearest neighbor spacing $a=0.25\lambda$. The different correlation functions are defined in Eqs.\ (\ref{eq:phiPhaseCorrelationFunc}) and (\ref{eq:dDOTd}) and indicated by the key in the figure. A grey line indicates a linear decrease from $+1$ to $-1$.   
}{fig:PhasorCorrelationFunctions}
\end{center}
\end{figure}

As we increase the atom number, the number of eigenmodes increases linearly, resulting in an even richer behavior. In Fig.\ \ref{fig:NAtomsChainModeVectors}, we plot a selection of the eigenvectors for a chain of $N=25$ atoms. We consider only those eigenvectors polarized perpendicular to the atomic axis ($\theta=\pi/2$). For each dipole in the chain we plot the amplitude and phase of the $\polvec_{\perp,1}$ polarized dipole vector as a phasor. 

Other than small deviations due to finite size effects, the general behavior of the eigenvectors has two main features. Firstly, as the mode index $\ell$ increases from $1$ to $N$, the average phase difference between nearest neighbor dipoles appears to increase. For $\ell=1$, the dipole oscillations are approximately all in phase with each other (indicated by the angle of their phasors). This is similar to modes $\vec{\mathbf{m}}_1$ and $\vec{\mathbf{m}}_{1'}$ from Fig.\ \ref{fig:ThreeAtomsExtinctionModes}. Conversely, for $\ell=N=25$, each dipole is approximately $\pi$ out of phase with its nearest neighbors, similar to $\vec{\mathbf{m}}_{3}$ and $\vec{\mathbf{m}}_{3'}$ in Fig.\ \ref{fig:ThreeAtomsExtinctionModes}. We can quantify this nearest neighbor phase difference by defining a phase correlation function
\begin{equation} \label{eq:phiPhaseCorrelationFunc}
  \langle \cos\varphi_{i,i+1}^{\ell} \rangle = \frac{1}{N-1}\sum_{i=1}^{N-1} \,\cos(\varphi_{i+1}^{\ell}-\varphi_{i}^{\ell}),
\end{equation}
where $\langle \rangle$ refers to averaging over every pair of nearest neighbor atoms. 
When all dipoles are in phase, the correlation function should be $\langle\cos\varphi_{i,i+1}^{\ell}\rangle\simeq 1$, and when all dipoles are out of phase, it should be $\langle\cos\varphi_{i,i+1}^{\ell}\rangle\simeq -1$. In Fig.\ \ref{fig:PhasorCorrelationFunctions} we plot $\langle\cos\varphi_{i,i+1}^{\ell}\rangle$ for increasing mode index and see an approximately linear decrease from $1$ to $-1$, confirming that the average phase difference between nearest neighbors does increase with increasing mode index.

In addition to the phase differences between neighboring dipoles, the amplitudes of the oscillating dipoles are not constant across the chain. Starting from $\ell=1$ and increasing the mode index, the amplitude envelopes can be described by harmonic modes of increasing order. The same is true starting from $\ell=N$ and decreasing mode index. In general, as $\ell$ tends towards $\ell=(N+1)/2$ from either direction, the amplitude envelope is a harmonic mode with $n$ antinodes, where $n=\ell$ for $\ell<(N+1)/2$ and $n=N+1-\ell$ for $\ell>(N+1)/2$. 

We can account for this change in amplitude of oscillation by defining a second nearest neighbor correlation function, 
\begin{align} \label{eq:dDOTd}
  \langle (\mathbf{d}_i^{\ell})^*\cdot\mathbf{d}_{i+1}^{\ell} \rangle =& \frac{1}{N-1}\sum_{i=1}^{N-1}\,
  \frac{(\mathbf{d}_i^{\ell})^*\cdot\mathbf{d}_{i+1}^{\ell}}{|\alpha_0E_k|^2}
\nonumber \\
  =& \frac{1}{N-1}\sum_{i=1}^{N-1}\,
  \frac{d_i^{\ell} d_{i+1}^{\ell}}{|\alpha_0E_k|^2} \, \mathrm{e}^{\mathrm{i}(\varphi_{i+1}^{\ell}-\varphi_{i}^{\ell})},
\end{align}
where $\mathbf{d}_i^{\ell}=d_i^{\ell}\mathop{\mathrm{e}}^{\mathrm{i}\varphi_i^{\ell}}\polvec_i^{\ell}$ is the dipole vector corresponding to the $\ell$th eigenvector with magnitude $d_i^{\ell}$, phase $\varphi_i^{\ell}$, and polarization $\polvec_i^{\ell}$. Eq.\ (\ref{eq:dDOTd}) is effectively the normalized expectation value of the dot product between two neighboring dipoles. We plot the real and imaginary parts of this separately in Fig.\ \ref{fig:PhasorCorrelationFunctions}. The imaginary part is always zero, although the real part, like $\langle\cos\varphi_{i,i+1}^{\ell}\rangle$, decreases (now nonlinearly) from $1$ to $-1$ with increasing mode index.

In this paper we always consider odd $N$. For even $N$, the same patterns in eigenvectors and eigenvalues appear. Because of the symmetry, the mode $\vec{\mathbf{m}}_N$ is antisymmetric about the center of the lattice rather than symmetric, although it is still fully antiphased. 

%%%%
\subsection{Eigenvalues} \label{sec:N25Eigenvalues}

% eigenvalues for N=25
\begin{figure}[t!]
\begin{center}
\includegraphics[width=8.6cm]{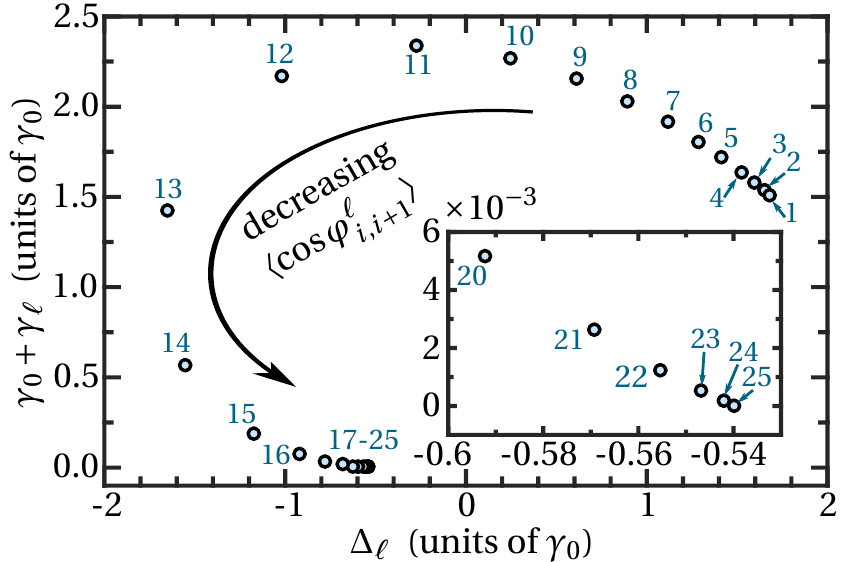}
\colorcaption
{Eigenvalues for a chain of $N=25$ atoms polarized perpendicular to the atomic axis ($\theta=\pi/2$) with nearest neighbour spacing $a=0.25\lambda$. The mode indices are labeled from $\ell=1$ to 25. The correlation function $\langle\cos\varphi_{i,i+1}^{\ell}\rangle$ is defined in Eq.\ (\ref{eq:phiPhaseCorrelationFunc}). }{fig:NAtomsChainEvals}
\end{center}
\end{figure}

% scattering cross section for N=25
\begin{figure}
\begin{center}
\includegraphics[width=8.6cm]{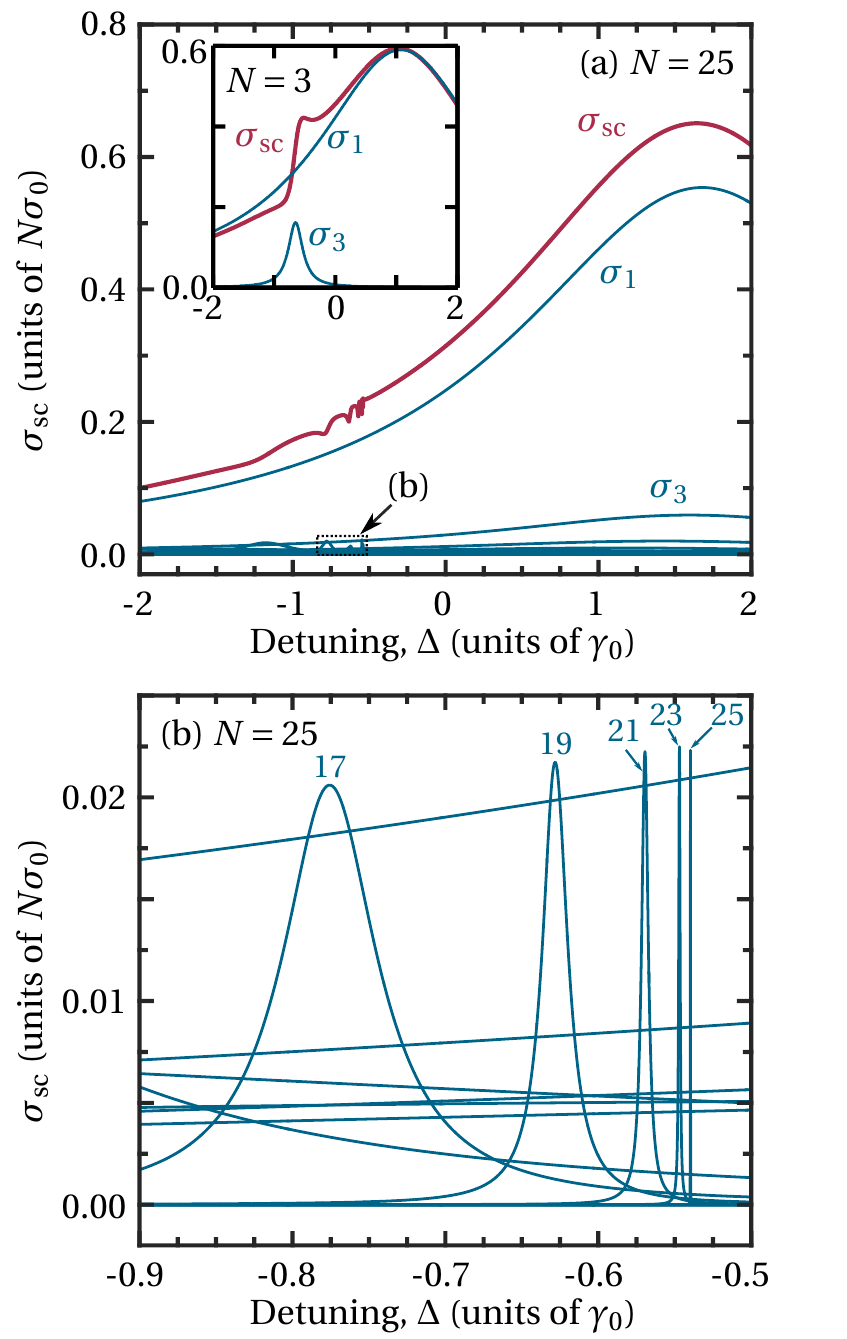}
\colorcaption
{Scattering cross section (red solid line) for a chain of $N=25$ (inset: $N=3$) atoms driven by a uniform driving field of varying detuning $\Delta$, propagating perpendicular to the atomic chain $\mathbf{k}_{\perp,2}$ and polarized perpendicular to the atomic chain $\polvec_{\perp,1}$. The atoms are separated by $a=0.25\lambda$. The individual contribution for each eigenmode is also plotted (blue lines), and the modes are labeled with the same mode indices as in Fig.\ \ref{fig:NAtomsChainModeVectors}, \ref{fig:PhasorCorrelationFunctions} and \ref{fig:NAtomsChainEvals}.
}{fig:NAtomsChainExtinctionCrossSectionModes}
\end{center}
\end{figure}

Our convention for mode index assignment has been such that the correlation functions continually decrease for increasing mode index. 
However, plotting the eigenvalues for an atomic separation of $a=0.25\lambda$ in Fig.\ \ref{fig:NAtomsChainEvals}, we see that the eigenvalues also depend on mode index, following a smooth arclike pathway through frequency space centered roughly on $(\Delta_{\ell}=0,\gamma_{\ell}=0$). The reason for this is that eigenvalues of each eigenmode are related to the total sum of each individual dipole vector (and also on the phase accumulated by scattering between dipoles). Similar eigenvalue plots have been made for 1D arrays \cite{Sutherland2016} as well as aperiodic Vogel spiral arrays \cite{Christofi2016} and random atomic ensembles \cite{Skipetrov2011,Bellando2014}. For random ensembles, the eigenvalue spectra typically consist of regions and narrow branches of randomly distributed eigenmodes.

In the Dicke picture \cite{Dicke1954,Gross1982}, an ensemble of $N$ atoms is confined to a volume much smaller than $\lambda$ in extent. In that situation, a mode like $\vec{\mathbf{m}}_1$, in which each dipole oscillates in phase, will behave like a macroscopic dipole, with a dipole moment $N$ times larger than each individual dipole moment. This coherent $N$-fold enhancement results in an enhanced scattering rate and a decay rate $N$ times larger than the decay rate of a single dipole. We can apply a similar idea to our chain of dipoles. Since the extent of the chain is now much larger than $\lambda$, the phase of the scattering between dipoles is also important, although we can still apply the idea of a coherent increase or decrease in the overall dipole moment of the ensemble. The overall dipole moment, and thus the eigenvalue of each eigenmode, are therefore clearly related to the relative phase and magnitude of each dipole in the chain. \MyHighlight{This will be discussed further in Sec.\ \ref{sec:EvalDependenceOnSpacing}.}

%%%%
\subsection{Scattering cross section}

Let us now consider which modes can be addressed by a uniform driving field 
\MyHighlight{with polarization and propagation wavevector both orthogonal to each other and to the atomic chain ($\polvec_{\perp,1},\mathbf{k}_{\perp,2}$).}
In Fig.\ \ref{fig:NAtomsChainExtinctionCrossSectionModes} we find that the scattering cross section is dominated by the fully in phase mode $\sigma_1$.  
For this atomic separation ($a=0.25\lambda$), the in phase mode is superradiant and blue shifted. In contrast to the three atom case in Fig.\ \ref{fig:ThreeAtomsExtinctionModes}, the higher index modes are now only very weakly coupled to the driving field. This is because the overlap between the uniform driving field and the out of phase dipoles is small. The perturbation of these highly subradiant modes is still visible in the total cross section, although in practice would likely be washed out by experimental uncertainties in the atomic position. Notice also that only the odd numbered modes are visible. This is because the even numbered modes are all antisymmetric whilst the odd numbered modes (like the uniform driving field) are symmetric. 

%%%%
\subsection{Eigenvalue dependence on atomic spacing} \label{sec:EvalDependenceOnSpacing}

% eigenvalues as a function of lattice spacing for N=25
\begin{figure*}[t]
\begin{center}
\includegraphics[width=15.5cm]{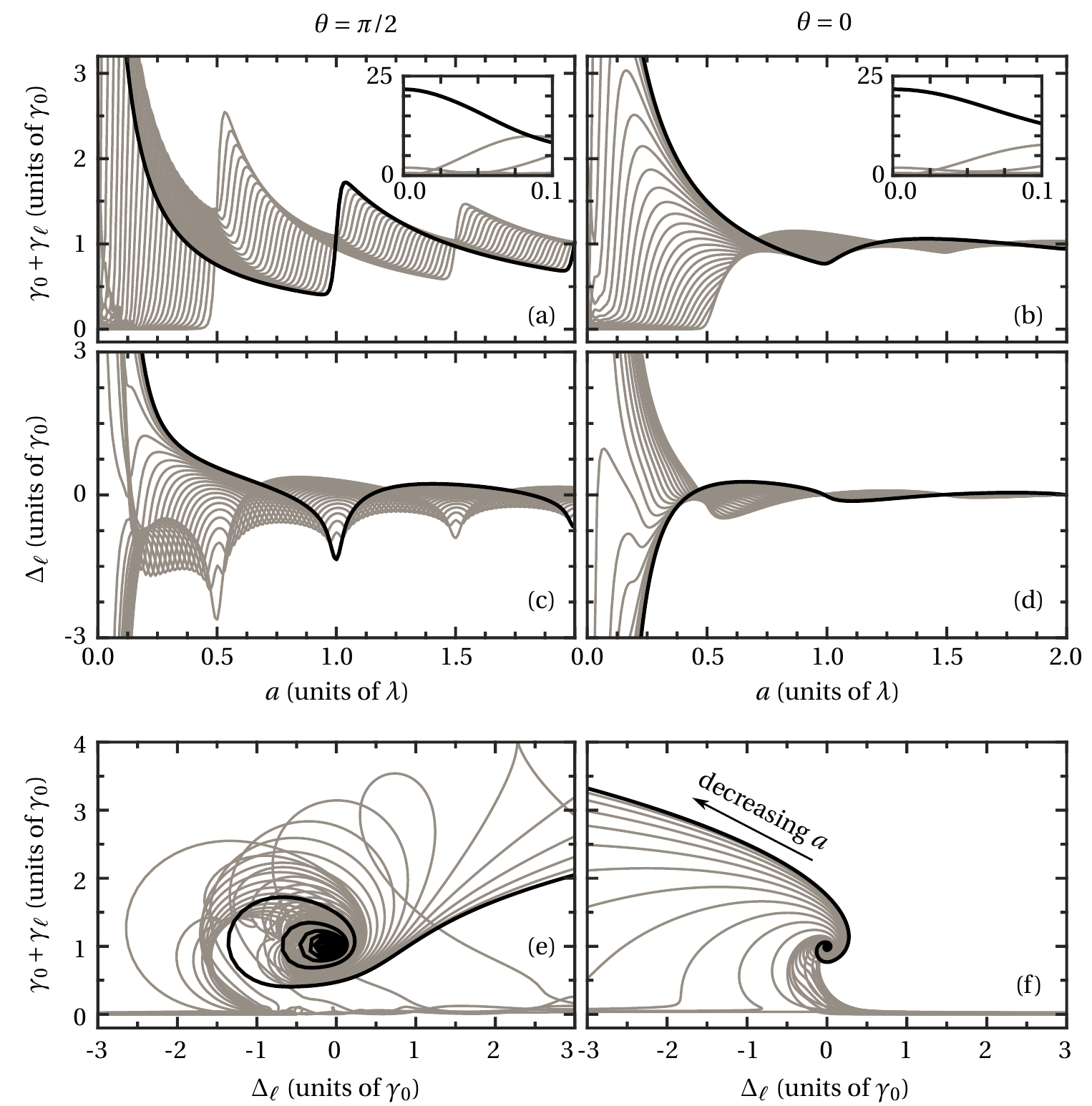}
\colorcaption{Eigenvalue dependence on atomic spacing for a chain of $N=25$ atoms. The angle between the atomic separation and the polarization vector is $\theta=\pi/2$ (a,c,e) and $\theta=0$ (b,d,f). In (a--d) the half-decay rates (a,b) and eigenvalue shifts (c,d) are plotted separately as a function of atom spacing. In (e,f) the same eigenvalues are plotted together, starting at the center with large atomic spacing and spiralling out with decreasing atomic spacing. The half-decay rates in (a) are similar to those in Figs.\ \ref{fig:NAtomChainShiftsWidthsInfSingle} and \ref{fig:NAtomChainShiftsWidthsInf}(a).
}{fig:NAtomChainEvalsSpiralPlot}
\end{center}
\end{figure*}

So far we have only considered a single atomic spacing, $a=0.25\lambda$. However, as the dipole--dipole interaction (\ref{eq:Vdd}) depends on atomic spacing, so will the eigenvalues (the correlation functions in Fig.\ \ref{fig:PhasorCorrelationFunctions} do not change significantly for different atomic spacings; individual mode vectors may have slightly different phases or amplitudes, but stay approximately the same as in Fig.\ \ref{fig:NAtomsChainModeVectors}). In Fig.\ \ref{fig:NAtomChainEvalsSpiralPlot} we plot the eigenvalues for the chain of $N=25$ atoms for $\theta=\pi/2$ and $\theta=0$ as a function of atomic spacing. We highlight the fully in phase mode $\vec{\mathbf{m}}_1$ with a black solid line. This mode tends to a decay rate of $\gamma_0+\gamma_1\simeq 22\gamma_0$ as $a\to 0$ for either orientation. As discussed in Sec.\ \ref{sec:N25Eigenvalues}, this is analogous to the Dicke fully symmetric state, which for $a\to 0 $ becomes $\gamma_0+\gamma_1=N\gamma_0$. However, because the mode has a nonuniform amplitude envelope such that the dipole moments are larger in the center and smaller at the edges, the fully in phase mode considered here does not completely reproduce the Dicke picture.

In Fig.\ \ref{fig:NAtomChainEvalsSpiralPlot}(e,f), we plot the eigenvalue spectra. As already noted, modes with superradiant decay rates and small $a$ are all blue shifted for $\theta=\pi/2$ and are all red shifted for $\theta=0$. This is because of the difference in sign of the dipole--dipole interaction (\ref{eq:Vdd}) between the two different orientations.

%%%%%%%%
\section{Atom chain, varying $N$} \label{sec:VaryingN}
%%%%
\subsection{Convergent and divergent eigenvalue limits} \label{sec:ConvDivSums}

% eigenvalues for N=11, N=25 and N=51
\begin{figure*}
\begin{center}
\includegraphics[width=15.5cm]{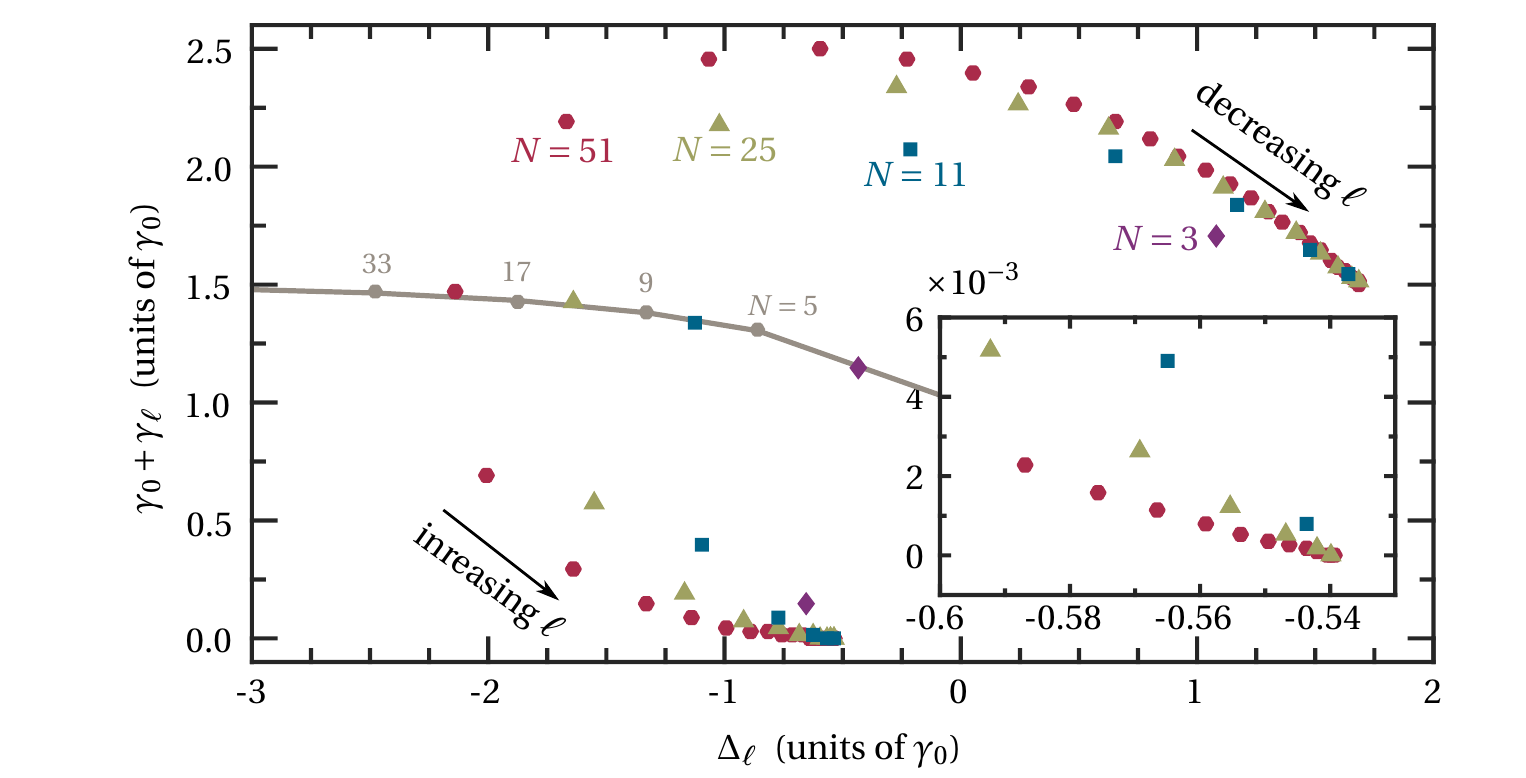}
\colorcaption
{Eigenvalues for a chain of atoms polarized perpendicular to the atomic axis ($\theta=\pi/2$) with nearest neighbour spacing $a=0.25\lambda$. The different markers correspond to atom numbers $N=3$ (purple diamonds), $N=11$ (blue squares), $N=25$ (green triangles), and $N=51$ (red hexagons). The grey line and markers plot the predicted eigenvalues (\ref{eq:DeltaGammaAnsatz}) assuming an eigenvector with nearest neighbor phase difference \MyHighlight{$\varphi^{\ell}_{j,j+1}=\pi/2$}. This is to demonstrate how the $\ell=(N+1)/2$ mode cooperative shifts do not converge as $N\to\infty$.  
}{fig:MultipleAtomChainsEvals}
\end{center}
\end{figure*}

% atom in a cavity plot
\begin{figure*}
\begin{center}
\includegraphics[width=15.5cm]{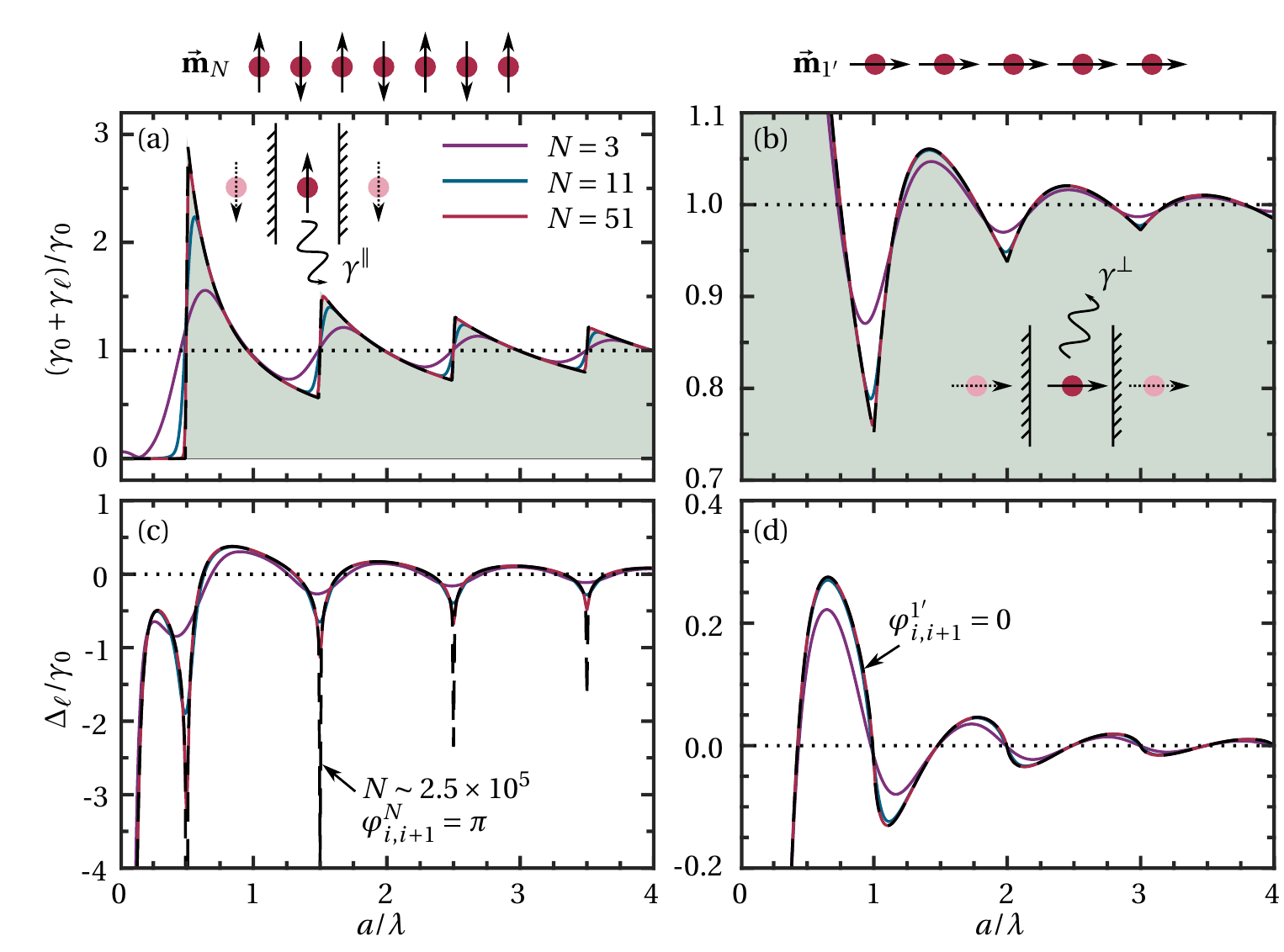}
\colorcaption
{(a,b) Half-decay rates and (c,d) cooperative shifts for chains of $N=3$ (purple), $N=11$ (blue), and $N=51$ (red) atoms. In (a,c), the atoms are polarized perpendicular to the atomic axis ($\theta=\pi/2$) and we consider mode $\vec{\mathbf{m}}_{N}$; in (b,d) the atoms are polarized parallel to the atomic axis ($\theta=0$) and we consider mode $\vec{\mathbf{m}}_{1'}$. The shaded areas plot the half-decay rates for a single atom between two mirrors, (a) polarized parallel to the mirrors [$\gamma_{\parallel}$, Eq.\ (\ref{eq:GammaPara})] and (b) polarized perpendicular to the mirrors [$\gamma_{\perp}$, Eq.\ (\ref{eq:GammaPerp})]. We also plot with black dotted lines the half-decay rates and shifts calculated using the eigenvector ansatz in Eq.\ (\ref{eq:DeltaGammaAnsatz}), with nearest neighbor phase difference $\varphi_{i,i+1}^{\ell}=\pi$ (a,c) and $\varphi_{i,i+1}^{\ell}=0$ (b,d), assuming an atom number of $N=(2.5\times 10^5)+1$.} {fig:NAtomChainShiftsWidthsInf}
\end{center}
\end{figure*}

Increasing the atom number from $N=3$ to $N=25$, we see an increase in the number of eigenmodes as well as the complexity of their behavior. In Fig.\ \ref{fig:MultipleAtomChainsEvals}, we plot the eigenvalues for an atomic spacing of $a=0.25\lambda$ as we did in Fig.\ \ref{fig:NAtomsChainEvals}, but now for different numbers of atoms ($N=3, 11, 25, 51$). We notice firstly that the maximal modes ($\ell\simeq1$, $\ell\simeq N$) at either end of the arcs of eigenvalues
\MyHighlight{appear to converge to limiting values as $N\to\infty$.} 
 However, for the modes around $\ell\simeq N/2$, whilst for any given value of $N$ the eigenvalues are well defined and finite, as we increase $N$, the eigenvalues do not appear to converge as they did for $\ell\simeq1$ and $\ell\simeq N$.

%%%%%%%%%%%%%% NEW %%%%%%%%%%%%%%%%%
\MyHighlight{ 
Looking at the eigenvectors in Fig.\ \ref{fig:NAtomsChainModeVectors}, it is possible to make guesses as to the general eigenvector behavior for the eigenvectors of a chain of $N$ atoms. Let us consider the modes $\vec{\mathbf{m}}_1$ and $\vec{\mathbf{m}}_{(N+1)/2}$. For these modes, the $j$th dipole has the general form $\mathbf{d}_{j}^{\ell}\simeq d\cos(j\varphi^{\ell})\,\polvec$ \footnote{This form for $\mathbf{d}_j^{\ell}$ is a good approximation for some, but not all, of the eigenvectors}, where the nearest neighbor phase difference $\varphi^{\ell}_{j,j+1} = \varphi^{\ell}$ is the same for all ($j,j+1$). We are interested in the limit $N\to\infty$ and so we ignore the edge effects and amplitude envelopes. Substituting $\mathbf{d}_{j}^{\ell}$ into Eq.\ (\ref{eq:coupledDipoleEq}) results in eigenvalues of the form \cite{Nienhuis1987,Kramer,Markel1993,Roof2016} 
\begin{subequations}\label{eq:DeltaGammaAnsatz}\begin{align}
  \Delta_{\ell} &= -\alpha_0\gamma_0 \sum_{j=-\infty}^{\infty} \mathop{\mathrm{Re}}\left[\cos(j\varphi^{\ell})\mathsf{G}_{0j}\right] (1-\delta_{0j}), \label{eq:DeltaAnsatz}\\
  \gamma_{\ell} &=\alpha_0\gamma_0 \sum_{j=-\infty}^{\infty} \mathop{\mathrm{Im}}\left[\cos(j\varphi^{\ell})\mathsf{G}_{0j}\right](1-\delta_{0j}).
\end{align}\end{subequations} 
If the position of the $j$th dipole is $aj$, then $\mathsf{G}_{0j}$ is proportional to $1/|j|$, $1/|j|^2$, and $1/|j|^3$. The sum over a series $\sum_{j=1}^\infty 1/j^{\rho}$ is \textit{absolutely} convergent if $\rho>1$. However, if $\rho=1$ then the sum is convergent only if the sign of the numerator alternates (with some periodicity); $\sum_{j=1}^\infty 1/j$ does not converge. Such a sum is \textit{conditionally} convergent.   
}

For $\theta=0$, the $1/j$ terms cancel in $\mathsf{G}_{0j}$ (\ref{eq:dipoleField}), meaning the eigenvalues are always absolutely convergent since $\mathsf{G}_{0j}$ only depends on $1/j^2$ and $1/j^3$. For $\theta=\pi/2$, however, the $1/j$ terms do not cancel, meaning the eigenvalues from (\ref{eq:DeltaGammaAnsatz}) become (ignoring the $1/j^2$ and $1/j^3$ terms)
\begin{subequations} \begin{align} 
  \Delta_{\ell}&\propto\sum_{j=1}^{\infty} \frac{\cos(j\varphi^{\ell})\cos(kaj)}{j}, \label{eq:DeltaSum} \\ 
  \gamma_{\ell}&\propto\sum_{j=1}^{\infty} \frac{\cos(j\varphi^{\ell})\sin(kaj)}{j}. \label{eq:gammaSum}
\end{align}\end{subequations}
The numerator in (\ref{eq:gammaSum}) changes sign as a function of $j$ and so $\gamma_{\ell}$ is always convergent. However, depending on the relationship between $\varphi^{\ell}$ and $aj$, the numerator in (\ref{eq:DeltaSum}) may or may not have an alternating sign. For example, mode $\ell=1$ in Fig.\ \ref{fig:MultipleAtomChainsEvals} with atom spacing $a=\lambda/4$ can be described with a phase difference $\varphi^{\ell}=0$. In this case, the numerator in (\ref{eq:DeltaSum}) is $\cos(j\pi/2)$ which changes sign as a function of $j$ and therefore results in a converging series. This is confirmed by our observation in Fig.\ \ref{fig:MultipleAtomChainsEvals} that the eigenvalues for $\ell=1$ appear to converge as $N$ increases. Conversely, for $\ell=(N+1)/2$, $\varphi^{\ell}=\pi/2$, and so the numerator in (\ref{eq:DeltaSum}) becomes \MyHighlight{$\cos^2(j\pi/2)$} which always has the same sign, and therefore the shifts $\Delta_{(N+1)/2}$ do not converge as $N\to\infty$.  

Similar discussions of the convergence and divergence of the eigenvalues of a 1D chain of dipoles, as well as analytic solutions in various limits, can be found in \cite{Nienhuis1987,Markel1993,Kramer}.

%%%%
\subsection{Atom in a cavity}

In Fig.\ \ref{fig:NAtomChainShiftsWidthsInfSingle}, we showed that the decay rate of a single atom inside a cavity polarized parallel to the cavity mirrors is approximated well by the decay rate of the antiphased eigenmode $\vec{\mathbf{m}}_{N}$ of a chain of $N=51$ atoms polarized perpendicularly to the atomic axis. In Fig.\ \ref{fig:NAtomChainShiftsWidthsInf}, we plot the decay rates and also the energy shifts as a function of lattice spacing for $N=\{3,11,51\}$. In Fig.\ \ref{fig:NAtomChainShiftsWidthsInf}(a) we find that the decay rates of the atom chains in the antiphased mode $\vec{\mathbf{m}}_N$ tend towards the decay rate of a single atom in a cavity polarized parallel to the cavity mirrors. In Fig.\ \ref{fig:NAtomChainShiftsWidthsInf}(b) we find the same is true when considering the fully in phase mode $\vec{\mathbf{m}}_{1'}$ and an atom polarized perpendicular to the mirrors. In Fig.\ \ref{fig:NAtomChainShiftsWidthsInf}(c,d) we plot the cooperative shifts for these same modes. Comparing these with the shifts calculated using Eq.\ (\ref{eq:DeltaAnsatz}), we see that as predicted in Sec.\ \ref{sec:ConvDivSums}, the shifts diverge logarithmically when \MyHighlight{$a=Z_{\text{odd}}\lambda/2$} for odd integers \MyHighlight{$Z_{\text{odd}}$} and $\theta=\pi/2$ (c) \footnote{The finite shifts of the black dashed lines in Fig.\ \ref{fig:NAtomChainShiftsWidthsInf}(c) at $a=2.5\lambda$ and $a=3.5\lambda$ are simply the computational limit rather than any physical limit.}, otherwise the shifts (and widths) converge for all other $a$ for both $\theta=\pi/2$ and $\theta=0$.

%%%%%%%%
\section{Comparison between 1D and 2D arrays} \label{sec:ComparisonBetween1Dand2D}

\MyHighlight{Many of the features we observe in this paper for 1D atomic arrays are similar to the behaviors that have been observed in previous studies of 2D atomic arrays \cite{Bettles2015d,Bettles2016,Kramer,Jenkins2012}. For example, in Figs.\ 2 and 3 of \cite{Bettles2015d}, the cross section lineshapes for 2D square and kagome arrays can exhibit Fano-like resonances due to interferences between multiple cooperative eigenmodes, similar to the lineshapes observed in Figs.\ \ref{fig:ThreeAtomsExtinctionModes} and \ref{fig:NAtomsChainExtinctionCrossSectionModes}. The behaviors of these eigenvalues and eigenvectors also exhibit similarities. For example, Fig.\ 1 of \cite{Bettles2015d} and Fig.\ 3 of \cite{Kramer} show a similar dependence of the eigenvalues on nearest neighbor spacing to that observed for a 1D chain in Fig.\ \ref{fig:NAtomChainEvalsSpiralPlot}. However, a crucial difference is that in 1D the spacing between pairs of atoms is commensurate, i.e., it is always an integer multiple of $a$. In 2D, however, the atom spacings are incommensurate, since next-nearest neighbors are separated by $\sqrt{2}a$ and so on for next-next-nearest neighbors etc. This means that whilst the eigenvalue resonances and poles in 1D (Fig.\ \ref{fig:NAtomChainEvalsSpiralPlot}) occur at half-integer multiples of $a=\lambda/2$, the equivalent resonances in 2D do not occur at such regular intervals (c.f.\ the peaks and troughs of the transmission in Fig.\ 2 of \cite{Bettles2016}, as well as Fig.\ 3 in \cite{Kramer}). One other point of comparison is in the form of the eigenvectors. For 1D chains, the eigenvectors form well defined patterns, ranging from all dipoles oscillating in phase to all oscillating out of phase with their nearest neighbors. In 2D with uniform driving, again the dominant eigenmode is typically one in which all dipoles oscillate in phase and are aligned along the polarization direction of the driving field (Fig.\ 3(b) of \cite{Bettles2015d}). The extra dimension however means that, in general, the structure of the eigenvectors in 2D is more complicated, as demonstrated by the hybrid mode in Fig.\ 3(c) of \cite{Bettles2015d}, exhibiting both in-phase and out-of-phase behavior, alternating between different rows of dipoles. }

\MyHighlight{
The underlying similarities between the cooperative behavior of 1D and 2D arrays mean that understanding the 1D system better should in turn provide insight into the more complicated behavior of 2D and higher dimensional configurations. For example, it may be possible to define similar phase correlation functions in 2D as for those defined in (\ref{eq:phiPhaseCorrelationFunc}) and (\ref{eq:dDOTd}), thus potentially finding patterns or structures in the otherwise complicated eigenvector behaviors. 
}

%%%%%%%%
\section{Conclusions and Outlook} \label{sec:Conclusions}
In conclusion, we have investigated the cooperative behavior of 1D atomic ensembles in free space, calculating the scattering cross sections and how these can be explained by considering the eigenmodes of the system. The complex symmetry of the coupling between the dipoles results in nonorthogonal eigenvectors which interfere, producing striking asymmetric Fano resonances in the scattering. The eigenvalues are also complex, meaning each eigenmode experiences an energy shift as well as a broadening or narrowing of the linewidth, corresponding to a modification of the scattering or decay rate. Even for just three atoms in a line, a broad range of cooperative behaviors are accessible, including strong superradiance, subradiance, line shifts and mode interferences, tunable by simply changing the driving field polarization and direction. Analyzing the eigenvectors of a chain of $N=25$ atoms we find the eigenvectors range from the dipoles all oscillating in phase to all oscillating out of phase with their nearest neighbors. This eigenvector behavior relates to the eigenvalues as well. For increasing atom number, some eigenvalues diverge whilst others converge to a behavior described by a single atom between two mirrors, demonstrating an analog between dipolar interactions and cavity QED. 
 
\MyHighlight{The classical model described in Sec.\ \ref{sec:CoupledClassicalDipoles} is a good approximation to the full quantum model, provided the amplitude of the driving field is sufficiently weak, 
$E_k\ll |\mathbf{d}_{\nu g}|/\alpha_0$. For stronger driving, finite excited state populations result in nonlinear saturation effects in the cross section, for example attenuating some of the narrower weaker eigenmodes and modifying the overall cross section lineshapes. This has been the subject of recent work \cite{Kramer2015,Lee2016} and will be investigated further in the future.
}
Experimental limitations such as imperfect trapping localization and finite filling factors may also affect the cooperative behavior discussed in this paper \MyHighlight{(e.g., by causing the narrow resonances to wash out)}, and so would need to be accounted for, as was done in \cite{Bettles2015d,Bettles2016,Jenkins2012}. 
The methods presented in this paper can be applied to many different configurations, not just of atomic dipoles, but also quantum dots, metamolecules, nanoresonators etc.\ We hope our study into the interesting resonant behavior of 1D systems will inspire further investigation and help to begin to explain the intricate mode behaviors observed in higher dimensional systems \cite{Jenkins2012,Bettles2015d,Bettles2016}.

The data presented  in  this  paper   can be found in Ref.\ \footnote{DOI: \href{http://dx.doi.org/10.15128/r2gx41mh849}{10.15128/r2gx41mh849}}

%%%%%%%%
\begin{acknowledgements}
We thank H.\ Ritsch, C.\ Genes, S.\ Kr\"{a}mer, B.\ Hopkins, R.T.\ Sutherland, and J.\ Ruostekoski for helpful discussions. 
We acknowledge funding from the UK EPSRC (Grant No.\ EP/L023024/1). 
\end{acknowledgements}

%%%%%%%% bibliography
\bibliography{library}

\begin{thebibliography}{87}%
\makeatletter
\providecommand \@ifxundefined [1]{%
 \@ifx{#1\undefined}
}%
\providecommand \@ifnum [1]{%
 \ifnum #1\expandafter \@firstoftwo
 \else \expandafter \@secondoftwo
 \fi
}%
\providecommand \@ifx [1]{%
 \ifx #1\expandafter \@firstoftwo
 \else \expandafter \@secondoftwo
 \fi
}%
\providecommand \natexlab [1]{#1}%
\providecommand \enquote  [1]{``#1''}%
\providecommand \bibnamefont  [1]{#1}%
\providecommand \bibfnamefont [1]{#1}%
\providecommand \citenamefont [1]{#1}%
\providecommand \href@noop [0]{\@secondoftwo}%
\providecommand \href [0]{\begingroup \@sanitize@url \@href}%
\providecommand \@href[1]{\@@startlink{#1}\@@href}%
\providecommand \@@href[1]{\endgroup#1\@@endlink}%
\providecommand \@sanitize@url [0]{\catcode `\\12\catcode `\$12\catcode
  `\&12\catcode `\#12\catcode `\^12\catcode `\_12\catcode `\%12\relax}%
\providecommand \@@startlink[1]{}%
\providecommand \@@endlink[0]{}%
\providecommand \url  [0]{\begingroup\@sanitize@url \@url }%
\providecommand \@url [1]{\endgroup\@href {#1}{\urlprefix }}%
\providecommand \urlprefix  [0]{URL }%
\providecommand \Eprint [0]{\href }%
\providecommand \doibase [0]{http://dx.doi.org/}%
\providecommand \selectlanguage [0]{\@gobble}%
\providecommand \bibinfo  [0]{\@secondoftwo}%
\providecommand \bibfield  [0]{\@secondoftwo}%
\providecommand \translation [1]{[#1]}%
\providecommand \BibitemOpen [0]{}%
\providecommand \bibitemStop [0]{}%
\providecommand \bibitemNoStop [0]{.\EOS\space}%
\providecommand \EOS [0]{\spacefactor3000\relax}%
\providecommand \BibitemShut  [1]{\csname bibitem#1\endcsname}%
\let\auto@bib@innerbib\@empty
%</preamble>
\bibitem [{\citenamefont {Dicke}(1954)}]{Dicke1954}%
  \BibitemOpen
  \bibfield  {author} {\bibinfo {author} {\bibfnamefont {R.}~\bibnamefont
  {Dicke}},\ }\href {\doibase 10.1103/PhysRev.93.99} {\bibfield  {journal}
  {\bibinfo  {journal} {Phys. Rev.}\ }\textbf {\bibinfo {volume} {93}},\
  \bibinfo {pages} {99} (\bibinfo {year} {1954})}\BibitemShut {NoStop}%
\bibitem [{\citenamefont {Stephen}(1964)}]{Stephen1964}%
  \BibitemOpen
  \bibfield  {author} {\bibinfo {author} {\bibfnamefont {M.~J.}\ \bibnamefont
  {Stephen}},\ }\href {\doibase 10.1063/1.1725188} {\bibfield  {journal}
  {\bibinfo  {journal} {J. Chem. Phys.}\ }\textbf {\bibinfo {volume} {40}},\
  \bibinfo {pages} {669} (\bibinfo {year} {1964})}\BibitemShut {NoStop}%
\bibitem [{\citenamefont {Lehmberg}(1970)}]{Lehmberg1970}%
  \BibitemOpen
  \bibfield  {author} {\bibinfo {author} {\bibfnamefont {R.~H.}\ \bibnamefont
  {Lehmberg}},\ }\href
  {http://onlinelibrary.wiley.com/doi/10.1002/cbdv.200490137/abstract
  http://pra.aps.org/abstract/PRA/v2/i3/p883{\_}1} {\bibfield  {journal}
  {\bibinfo  {journal} {Phys. Rev. A}\ }\textbf {\bibinfo {volume} {2}},\
  \bibinfo {pages} {883} (\bibinfo {year} {1970})}\BibitemShut {NoStop}%
\bibitem [{\citenamefont {Friedberg}\ \emph {et~al.}(1973)\citenamefont
  {Friedberg}, \citenamefont {Hartmann},\ and\ \citenamefont
  {Manassah}}]{Friedberg1973}%
  \BibitemOpen
  \bibfield  {author} {\bibinfo {author} {\bibfnamefont {R.}~\bibnamefont
  {Friedberg}}, \bibinfo {author} {\bibfnamefont {S.~R.}\ \bibnamefont
  {Hartmann}}, \ and\ \bibinfo {author} {\bibfnamefont {J.~T.}\ \bibnamefont
  {Manassah}},\ }\href {\doibase 10.1016/0370-1573(73)90001-X} {\bibfield
  {journal} {\bibinfo  {journal} {Phys. Rep.}\ }\textbf {\bibinfo {volume}
  {7}},\ \bibinfo {pages} {101} (\bibinfo {year} {1973})}\BibitemShut {NoStop}%
\bibitem [{\citenamefont {Gross}\ and\ \citenamefont
  {Haroche}(1982)}]{Gross1982}%
  \BibitemOpen
  \bibfield  {author} {\bibinfo {author} {\bibfnamefont {M.}~\bibnamefont
  {Gross}}\ and\ \bibinfo {author} {\bibfnamefont {S.}~\bibnamefont
  {Haroche}},\ }\href {\doibase 10.1016/0370-1573(82)90102-8} {\bibfield
  {journal} {\bibinfo  {journal} {Phys. Rep.}\ }\textbf {\bibinfo {volume}
  {93}},\ \bibinfo {pages} {301} (\bibinfo {year} {1982})}\BibitemShut
  {NoStop}%
\bibitem [{\citenamefont {DeVoe}\ and\ \citenamefont
  {Brewer}(1996)}]{DeVoe1996}%
  \BibitemOpen
  \bibfield  {author} {\bibinfo {author} {\bibfnamefont {R.~G.}\ \bibnamefont
  {DeVoe}}\ and\ \bibinfo {author} {\bibfnamefont {R.~G.}\ \bibnamefont
  {Brewer}},\ }\href {http://www.ncbi.nlm.nih.gov/pubmed/10060593} {\bibfield
  {journal} {\bibinfo  {journal} {Phys. Rev. Lett.}\ }\textbf {\bibinfo
  {volume} {76}},\ \bibinfo {pages} {2049} (\bibinfo {year}
  {1996})}\BibitemShut {NoStop}%
\bibitem [{\citenamefont {Inouye}\ \emph {et~al.}(1999)\citenamefont {Inouye},
  \citenamefont {Chikkatur}, \citenamefont {Stamper-Kurn}, \citenamefont
  {Stenger}, \citenamefont {Pritchard},\ and\ \citenamefont
  {Ketterle}}]{Inouye1999}%
  \BibitemOpen
  \bibfield  {author} {\bibinfo {author} {\bibfnamefont {S.}~\bibnamefont
  {Inouye}}, \bibinfo {author} {\bibfnamefont {A.~P.}\ \bibnamefont
  {Chikkatur}}, \bibinfo {author} {\bibfnamefont {D.~M.}\ \bibnamefont
  {Stamper-Kurn}}, \bibinfo {author} {\bibfnamefont {J.}~\bibnamefont
  {Stenger}}, \bibinfo {author} {\bibfnamefont {D.~E.}\ \bibnamefont
  {Pritchard}}, \ and\ \bibinfo {author} {\bibfnamefont {W.}~\bibnamefont
  {Ketterle}},\ }\href {http://www.sciencemag.org/content/285/5427/571.short}
  {\bibfield  {journal} {\bibinfo  {journal} {Science (80-. ).}\ }\textbf
  {\bibinfo {volume} {285}},\ \bibinfo {pages} {571} (\bibinfo {year}
  {1999})}\BibitemShut {NoStop}%
\bibitem [{\citenamefont {Yoshikawa}\ \emph {et~al.}(2005)\citenamefont
  {Yoshikawa}, \citenamefont {Torii},\ and\ \citenamefont
  {Kuga}}]{Yoshikawa2005}%
  \BibitemOpen
  \bibfield  {author} {\bibinfo {author} {\bibfnamefont {Y.}~\bibnamefont
  {Yoshikawa}}, \bibinfo {author} {\bibfnamefont {Y.}~\bibnamefont {Torii}}, \
  and\ \bibinfo {author} {\bibfnamefont {T.}~\bibnamefont {Kuga}},\ }\href
  {\doibase 10.1103/PhysRevLett.94.083602} {\bibfield  {journal} {\bibinfo
  {journal} {Phys. Rev. Lett.}\ }\textbf {\bibinfo {volume} {94}},\ \bibinfo
  {pages} {083602} (\bibinfo {year} {2005})}\BibitemShut {NoStop}%
\bibitem [{\citenamefont {Scheibner}\ \emph {et~al.}(2007)\citenamefont
  {Scheibner}, \citenamefont {Schmidt}, \citenamefont {Worschech},
  \citenamefont {Forchel}, \citenamefont {Bacher}, \citenamefont {Passow},\
  and\ \citenamefont {Hommel}}]{Scheibner2007}%
  \BibitemOpen
  \bibfield  {author} {\bibinfo {author} {\bibfnamefont {M.}~\bibnamefont
  {Scheibner}}, \bibinfo {author} {\bibfnamefont {T.}~\bibnamefont {Schmidt}},
  \bibinfo {author} {\bibfnamefont {L.}~\bibnamefont {Worschech}}, \bibinfo
  {author} {\bibfnamefont {A.}~\bibnamefont {Forchel}}, \bibinfo {author}
  {\bibfnamefont {G.}~\bibnamefont {Bacher}}, \bibinfo {author} {\bibfnamefont
  {T.}~\bibnamefont {Passow}}, \ and\ \bibinfo {author} {\bibfnamefont
  {D.}~\bibnamefont {Hommel}},\ }\href {\doibase 10.1038/nphys494} {\bibfield
  {journal} {\bibinfo  {journal} {Nat. Phys.}\ }\textbf {\bibinfo {volume}
  {3}},\ \bibinfo {pages} {106} (\bibinfo {year} {2007})}\BibitemShut {NoStop}%
\bibitem [{\citenamefont {Greenberg}\ and\ \citenamefont
  {Gauthier}(2012)}]{Greenberg2012}%
  \BibitemOpen
  \bibfield  {author} {\bibinfo {author} {\bibfnamefont {J.~A.}\ \bibnamefont
  {Greenberg}}\ and\ \bibinfo {author} {\bibfnamefont {D.~J.}\ \bibnamefont
  {Gauthier}},\ }\href {\doibase 10.1103/PhysRevA.86.013823} {\bibfield
  {journal} {\bibinfo  {journal} {Phys. Rev. A}\ }\textbf {\bibinfo {volume}
  {86}},\ \bibinfo {pages} {013823} (\bibinfo {year} {2012})}\BibitemShut
  {NoStop}%
\bibitem [{\citenamefont {Goban}\ \emph {et~al.}(2015)\citenamefont {Goban},
  \citenamefont {Hung}, \citenamefont {Hood}, \citenamefont {Yu}, \citenamefont
  {Muniz}, \citenamefont {Painter},\ and\ \citenamefont {Kimble}}]{Goban2015a}%
  \BibitemOpen
  \bibfield  {author} {\bibinfo {author} {\bibfnamefont {A.}~\bibnamefont
  {Goban}}, \bibinfo {author} {\bibfnamefont {C.-L.}\ \bibnamefont {Hung}},
  \bibinfo {author} {\bibfnamefont {J.~D.}\ \bibnamefont {Hood}}, \bibinfo
  {author} {\bibfnamefont {S.-P.}\ \bibnamefont {Yu}}, \bibinfo {author}
  {\bibfnamefont {J.~A.}\ \bibnamefont {Muniz}}, \bibinfo {author}
  {\bibfnamefont {O.}~\bibnamefont {Painter}}, \ and\ \bibinfo {author}
  {\bibfnamefont {H.~J.}\ \bibnamefont {Kimble}},\ }\href {\doibase
  10.1103/PhysRevLett.115.063601} {\bibfield  {journal} {\bibinfo  {journal}
  {Phys. Rev. Lett.}\ }\textbf {\bibinfo {volume} {115}},\ \bibinfo {pages}
  {063601} (\bibinfo {year} {2015})}\BibitemShut {NoStop}%
\bibitem [{\citenamefont {Guerin}\ \emph {et~al.}(2016)\citenamefont {Guerin},
  \citenamefont {Ara{\'{u}}jo},\ and\ \citenamefont {Kaiser}}]{Guerin2016a}%
  \BibitemOpen
  \bibfield  {author} {\bibinfo {author} {\bibfnamefont {W.}~\bibnamefont
  {Guerin}}, \bibinfo {author} {\bibfnamefont {M.~O.}\ \bibnamefont
  {Ara{\'{u}}jo}}, \ and\ \bibinfo {author} {\bibfnamefont {R.}~\bibnamefont
  {Kaiser}},\ }\href {\doibase 10.1103/PhysRevLett.116.083601} {\bibfield
  {journal} {\bibinfo  {journal} {Phys. Rev. Lett.}\ }\textbf {\bibinfo
  {volume} {116}},\ \bibinfo {pages} {083601} (\bibinfo {year}
  {2016})}\BibitemShut {NoStop}%
\bibitem [{\citenamefont {Ara{\'{u}}jo}\ \emph {et~al.}(2016)\citenamefont
  {Ara{\'{u}}jo}, \citenamefont {Kre{\v{s}}i{\'{c}}}, \citenamefont {Kaiser},\
  and\ \citenamefont {Guerin}}]{Araujo2016}%
  \BibitemOpen
  \bibfield  {author} {\bibinfo {author} {\bibfnamefont {M.~O.}\ \bibnamefont
  {Ara{\'{u}}jo}}, \bibinfo {author} {\bibfnamefont {I.}~\bibnamefont
  {Kre{\v{s}}i{\'{c}}}}, \bibinfo {author} {\bibfnamefont {R.}~\bibnamefont
  {Kaiser}}, \ and\ \bibinfo {author} {\bibfnamefont {W.}~\bibnamefont
  {Guerin}},\ }\href {\doibase 10.1103/PhysRevLett.117.073002} {\bibfield
  {journal} {\bibinfo  {journal} {Phys. Rev. Lett.}\ }\textbf {\bibinfo
  {volume} {117}},\ \bibinfo {pages} {073002} (\bibinfo {year}
  {2016})}\BibitemShut {NoStop}%
\bibitem [{\citenamefont {Roof}\ \emph {et~al.}(2016)\citenamefont {Roof},
  \citenamefont {Kemp}, \citenamefont {Havey},\ and\ \citenamefont
  {Sokolov}}]{Roof2016}%
  \BibitemOpen
  \bibfield  {author} {\bibinfo {author} {\bibfnamefont {S.~J.}\ \bibnamefont
  {Roof}}, \bibinfo {author} {\bibfnamefont {K.~J.}\ \bibnamefont {Kemp}},
  \bibinfo {author} {\bibfnamefont {M.~D.}\ \bibnamefont {Havey}}, \ and\
  \bibinfo {author} {\bibfnamefont {I.~M.}\ \bibnamefont {Sokolov}},\ }\href
  {\doibase 10.1103/PhysRevLett.117.073003} {\bibfield  {journal} {\bibinfo
  {journal} {Phys. Rev. Lett.}\ }\textbf {\bibinfo {volume} {117}},\ \bibinfo
  {pages} {073003} (\bibinfo {year} {2016})}\BibitemShut {NoStop}%
\bibitem [{\citenamefont {Keaveney}\ \emph {et~al.}(2012)\citenamefont
  {Keaveney}, \citenamefont {Sargsyan}, \citenamefont {Krohn}, \citenamefont
  {Hughes}, \citenamefont {Sarkisyan},\ and\ \citenamefont
  {Adams}}]{Keaveney2012}%
  \BibitemOpen
  \bibfield  {author} {\bibinfo {author} {\bibfnamefont {J.}~\bibnamefont
  {Keaveney}}, \bibinfo {author} {\bibfnamefont {A.}~\bibnamefont {Sargsyan}},
  \bibinfo {author} {\bibfnamefont {U.}~\bibnamefont {Krohn}}, \bibinfo
  {author} {\bibfnamefont {I.~G.}\ \bibnamefont {Hughes}}, \bibinfo {author}
  {\bibfnamefont {D.}~\bibnamefont {Sarkisyan}}, \ and\ \bibinfo {author}
  {\bibfnamefont {C.~S.}\ \bibnamefont {Adams}},\ }\href {\doibase
  10.1103/PhysRevLett.108.173601} {\bibfield  {journal} {\bibinfo  {journal}
  {Phys. Rev. Lett.}\ }\textbf {\bibinfo {volume} {108}},\ \bibinfo {pages}
  {173601} (\bibinfo {year} {2012})}\BibitemShut {NoStop}%
\bibitem [{\citenamefont {R{\"{o}}hlsberger}\ \emph {et~al.}(2010)\citenamefont
  {R{\"{o}}hlsberger}, \citenamefont {Schlage}, \citenamefont {Sahoo},
  \citenamefont {Couet},\ and\ \citenamefont {R{\"{u}}ffer}}]{Rohlsberger2010}%
  \BibitemOpen
  \bibfield  {author} {\bibinfo {author} {\bibfnamefont {R.}~\bibnamefont
  {R{\"{o}}hlsberger}}, \bibinfo {author} {\bibfnamefont {K.}~\bibnamefont
  {Schlage}}, \bibinfo {author} {\bibfnamefont {B.}~\bibnamefont {Sahoo}},
  \bibinfo {author} {\bibfnamefont {S.}~\bibnamefont {Couet}}, \ and\ \bibinfo
  {author} {\bibfnamefont {R.}~\bibnamefont {R{\"{u}}ffer}},\ }\href {\doibase
  10.1126/science.1187770} {\bibfield  {journal} {\bibinfo  {journal} {Science
  (80-. ).}\ }\textbf {\bibinfo {volume} {328}},\ \bibinfo {pages} {1248}
  (\bibinfo {year} {2010})}\BibitemShut {NoStop}%
\bibitem [{\citenamefont {Meir}\ \emph {et~al.}(2014)\citenamefont {Meir},
  \citenamefont {Schwartz}, \citenamefont {Shahmoon}, \citenamefont {Oron},\
  and\ \citenamefont {Ozeri}}]{Meir2014}%
  \BibitemOpen
  \bibfield  {author} {\bibinfo {author} {\bibfnamefont {Z.}~\bibnamefont
  {Meir}}, \bibinfo {author} {\bibfnamefont {O.}~\bibnamefont {Schwartz}},
  \bibinfo {author} {\bibfnamefont {E.}~\bibnamefont {Shahmoon}}, \bibinfo
  {author} {\bibfnamefont {D.}~\bibnamefont {Oron}}, \ and\ \bibinfo {author}
  {\bibfnamefont {R.}~\bibnamefont {Ozeri}},\ }\href {\doibase
  10.1103/PhysRevLett.113.193002} {\bibfield  {journal} {\bibinfo  {journal}
  {Phys. Rev. Lett.}\ }\textbf {\bibinfo {volume} {113}},\ \bibinfo {pages}
  {193002} (\bibinfo {year} {2014})}\BibitemShut {NoStop}%
\bibitem [{\citenamefont {Javanainen}\ \emph {et~al.}(2014)\citenamefont
  {Javanainen}, \citenamefont {Ruostekoski}, \citenamefont {Li},\ and\
  \citenamefont {Yoo}}]{Javanainen2014}%
  \BibitemOpen
  \bibfield  {author} {\bibinfo {author} {\bibfnamefont {J.}~\bibnamefont
  {Javanainen}}, \bibinfo {author} {\bibfnamefont {J.}~\bibnamefont
  {Ruostekoski}}, \bibinfo {author} {\bibfnamefont {Y.}~\bibnamefont {Li}}, \
  and\ \bibinfo {author} {\bibfnamefont {S.~M.}\ \bibnamefont {Yoo}},\ }\href
  {\doibase 10.1103/PhysRevLett.112.113603} {\bibfield  {journal} {\bibinfo
  {journal} {Phys. Rev. Lett.}\ }\textbf {\bibinfo {volume} {112}},\ \bibinfo
  {pages} {113603} (\bibinfo {year} {2014})}\BibitemShut {NoStop}%
\bibitem [{\citenamefont {Jennewein}\ \emph {et~al.}(2016)\citenamefont
  {Jennewein}, \citenamefont {Besbes}, \citenamefont {Schilder}, \citenamefont
  {Jenkins}, \citenamefont {Sauvan}, \citenamefont {Ruostekoski}, \citenamefont
  {Greffet}, \citenamefont {Sortais},\ and\ \citenamefont
  {Browaeys}}]{Jennewein2015d}%
  \BibitemOpen
  \bibfield  {author} {\bibinfo {author} {\bibfnamefont {S.}~\bibnamefont
  {Jennewein}}, \bibinfo {author} {\bibfnamefont {M.}~\bibnamefont {Besbes}},
  \bibinfo {author} {\bibfnamefont {N.~J.}\ \bibnamefont {Schilder}}, \bibinfo
  {author} {\bibfnamefont {S.~D.}\ \bibnamefont {Jenkins}}, \bibinfo {author}
  {\bibfnamefont {C.}~\bibnamefont {Sauvan}}, \bibinfo {author} {\bibfnamefont
  {J.}~\bibnamefont {Ruostekoski}}, \bibinfo {author} {\bibfnamefont {J.-J.}\
  \bibnamefont {Greffet}}, \bibinfo {author} {\bibfnamefont {Y.~R.~P.}\
  \bibnamefont {Sortais}}, \ and\ \bibinfo {author} {\bibfnamefont
  {A.}~\bibnamefont {Browaeys}},\ }\href {\doibase
  10.1103/PhysRevLett.116.233601} {\bibfield  {journal} {\bibinfo  {journal}
  {Phys. Rev. Lett.}\ }\textbf {\bibinfo {volume} {116}},\ \bibinfo {pages}
  {233601} (\bibinfo {year} {2016})}\BibitemShut {NoStop}%
\bibitem [{\citenamefont {Jenkins}\ \emph {et~al.}(2016)\citenamefont
  {Jenkins}, \citenamefont {Ruostekoski}, \citenamefont {Javanainen},
  \citenamefont {Bourgain}, \citenamefont {Jennewein}, \citenamefont
  {Sortais},\ and\ \citenamefont {Browaeys}}]{Jenkins2016}%
  \BibitemOpen
  \bibfield  {author} {\bibinfo {author} {\bibfnamefont {S.~D.}\ \bibnamefont
  {Jenkins}}, \bibinfo {author} {\bibfnamefont {J.}~\bibnamefont
  {Ruostekoski}}, \bibinfo {author} {\bibfnamefont {J.}~\bibnamefont
  {Javanainen}}, \bibinfo {author} {\bibfnamefont {R.}~\bibnamefont
  {Bourgain}}, \bibinfo {author} {\bibfnamefont {S.}~\bibnamefont {Jennewein}},
  \bibinfo {author} {\bibfnamefont {Y.~R.~P.}\ \bibnamefont {Sortais}}, \ and\
  \bibinfo {author} {\bibfnamefont {A.}~\bibnamefont {Browaeys}},\ }\href
  {\doibase 10.1103/PhysRevLett.116.183601} {\bibfield  {journal} {\bibinfo
  {journal} {Phys. Rev. Lett.}\ }\textbf {\bibinfo {volume} {116}},\ \bibinfo
  {pages} {183601} (\bibinfo {year} {2016})}\BibitemShut {NoStop}%
\bibitem [{\citenamefont {Rouabah}\ \emph {et~al.}(2014)\citenamefont
  {Rouabah}, \citenamefont {Samoylova}, \citenamefont {Bachelard},
  \citenamefont {Courteille}, \citenamefont {Kaiser},\ and\ \citenamefont
  {Piovella}}]{Rouabah2014}%
  \BibitemOpen
  \bibfield  {author} {\bibinfo {author} {\bibfnamefont {M.~T.}\ \bibnamefont
  {Rouabah}}, \bibinfo {author} {\bibfnamefont {M.}~\bibnamefont {Samoylova}},
  \bibinfo {author} {\bibfnamefont {R.}~\bibnamefont {Bachelard}}, \bibinfo
  {author} {\bibfnamefont {P.~W.}\ \bibnamefont {Courteille}}, \bibinfo
  {author} {\bibfnamefont {R.}~\bibnamefont {Kaiser}}, \ and\ \bibinfo {author}
  {\bibfnamefont {N.}~\bibnamefont {Piovella}},\ }\href
  {http://www.ncbi.nlm.nih.gov/pubmed/24979635} {\bibfield  {journal} {\bibinfo
   {journal} {J. Opt. Soc. Am. A}\ }\textbf {\bibinfo {volume} {31}},\ \bibinfo
  {pages} {1031} (\bibinfo {year} {2014})}\BibitemShut {NoStop}%
\bibitem [{\citenamefont {Oppel}\ \emph {et~al.}(2014)\citenamefont {Oppel},
  \citenamefont {Wiegner}, \citenamefont {Agarwal},\ and\ \citenamefont {von
  Zanthier}}]{Oppel2014}%
  \BibitemOpen
  \bibfield  {author} {\bibinfo {author} {\bibfnamefont {S.}~\bibnamefont
  {Oppel}}, \bibinfo {author} {\bibfnamefont {R.}~\bibnamefont {Wiegner}},
  \bibinfo {author} {\bibfnamefont {G.~S.}\ \bibnamefont {Agarwal}}, \ and\
  \bibinfo {author} {\bibfnamefont {J.}~\bibnamefont {von Zanthier}},\ }\href
  {\doibase 10.1103/PhysRevLett.113.263606} {\bibfield  {journal} {\bibinfo
  {journal} {Phys. Rev. Lett.}\ }\textbf {\bibinfo {volume} {113}},\ \bibinfo
  {pages} {263606} (\bibinfo {year} {2014})}\BibitemShut {NoStop}%
\bibitem [{\citenamefont {Luk'yanchuk}\ \emph {et~al.}(2010)\citenamefont
  {Luk'yanchuk}, \citenamefont {Zheludev}, \citenamefont {Maier}, \citenamefont
  {Halas}, \citenamefont {Nordlander}, \citenamefont {Giessen},\ and\
  \citenamefont {Chong}}]{Luk'yanchuk2010}%
  \BibitemOpen
  \bibfield  {author} {\bibinfo {author} {\bibfnamefont {B.}~\bibnamefont
  {Luk'yanchuk}}, \bibinfo {author} {\bibfnamefont {N.~I.}\ \bibnamefont
  {Zheludev}}, \bibinfo {author} {\bibfnamefont {S.~A.}\ \bibnamefont {Maier}},
  \bibinfo {author} {\bibfnamefont {N.~J.}\ \bibnamefont {Halas}}, \bibinfo
  {author} {\bibfnamefont {P.}~\bibnamefont {Nordlander}}, \bibinfo {author}
  {\bibfnamefont {H.}~\bibnamefont {Giessen}}, \ and\ \bibinfo {author}
  {\bibfnamefont {C.~T.}\ \bibnamefont {Chong}},\ }\href {\doibase
  10.1038/nmat2810} {\bibfield  {journal} {\bibinfo  {journal} {Nat. Mater.}\
  }\textbf {\bibinfo {volume} {9}},\ \bibinfo {pages} {707} (\bibinfo {year}
  {2010})}\BibitemShut {NoStop}%
\bibitem [{\citenamefont {Ghenuche}\ \emph {et~al.}(2012)\citenamefont
  {Ghenuche}, \citenamefont {Vincent}, \citenamefont {Laroche}, \citenamefont
  {Bardou}, \citenamefont {Ha{\"{i}}dar}, \citenamefont {Pelouard},\ and\
  \citenamefont {Collin}}]{Ghenuche2012}%
  \BibitemOpen
  \bibfield  {author} {\bibinfo {author} {\bibfnamefont {P.}~\bibnamefont
  {Ghenuche}}, \bibinfo {author} {\bibfnamefont {G.}~\bibnamefont {Vincent}},
  \bibinfo {author} {\bibfnamefont {M.}~\bibnamefont {Laroche}}, \bibinfo
  {author} {\bibfnamefont {N.}~\bibnamefont {Bardou}}, \bibinfo {author}
  {\bibfnamefont {R.}~\bibnamefont {Ha{\"{i}}dar}}, \bibinfo {author}
  {\bibfnamefont {J.-L.}\ \bibnamefont {Pelouard}}, \ and\ \bibinfo {author}
  {\bibfnamefont {S.}~\bibnamefont {Collin}},\ }\href {\doibase
  10.1103/PhysRevLett.109.143903} {\bibfield  {journal} {\bibinfo  {journal}
  {Phys. Rev. Lett.}\ }\textbf {\bibinfo {volume} {109}},\ \bibinfo {pages}
  {143903} (\bibinfo {year} {2012})}\BibitemShut {NoStop}%
\bibitem [{\citenamefont {Jenkins}\ and\ \citenamefont
  {Ruostekoski}(2013)}]{Jenkins2013}%
  \BibitemOpen
  \bibfield  {author} {\bibinfo {author} {\bibfnamefont {S.~D.}\ \bibnamefont
  {Jenkins}}\ and\ \bibinfo {author} {\bibfnamefont {J.}~\bibnamefont
  {Ruostekoski}},\ }\href {\doibase 10.1103/PhysRevLett.111.147401} {\bibfield
  {journal} {\bibinfo  {journal} {Phys. Rev. Lett.}\ }\textbf {\bibinfo
  {volume} {111}},\ \bibinfo {pages} {147401} (\bibinfo {year}
  {2013})}\BibitemShut {NoStop}%
\bibitem [{\citenamefont {Hopkins}\ \emph {et~al.}(2013)\citenamefont
  {Hopkins}, \citenamefont {Poddubny}, \citenamefont {Miroshnichenko},\ and\
  \citenamefont {Kivshar}}]{Hopkins2013}%
  \BibitemOpen
  \bibfield  {author} {\bibinfo {author} {\bibfnamefont {B.}~\bibnamefont
  {Hopkins}}, \bibinfo {author} {\bibfnamefont {A.~N.}\ \bibnamefont
  {Poddubny}}, \bibinfo {author} {\bibfnamefont {A.~E.}\ \bibnamefont
  {Miroshnichenko}}, \ and\ \bibinfo {author} {\bibfnamefont {Y.~S.}\
  \bibnamefont {Kivshar}},\ }\href {\doibase 10.1103/PhysRevA.88.053819}
  {\bibfield  {journal} {\bibinfo  {journal} {Phys. Rev. A}\ }\textbf {\bibinfo
  {volume} {88}},\ \bibinfo {pages} {053819} (\bibinfo {year}
  {2013})}\BibitemShut {NoStop}%
\bibitem [{\citenamefont {Puthumpally-Joseph}\ \emph
  {et~al.}(2014)\citenamefont {Puthumpally-Joseph}, \citenamefont {Sukharev},
  \citenamefont {Atabek},\ and\ \citenamefont
  {Charron}}]{Puthumpally-Joseph2014a}%
  \BibitemOpen
  \bibfield  {author} {\bibinfo {author} {\bibfnamefont {R.}~\bibnamefont
  {Puthumpally-Joseph}}, \bibinfo {author} {\bibfnamefont {M.}~\bibnamefont
  {Sukharev}}, \bibinfo {author} {\bibfnamefont {O.}~\bibnamefont {Atabek}}, \
  and\ \bibinfo {author} {\bibfnamefont {E.}~\bibnamefont {Charron}},\ }\href
  {\doibase 10.1103/PhysRevLett.113.163603} {\bibfield  {journal} {\bibinfo
  {journal} {Phys. Rev. Lett.}\ }\textbf {\bibinfo {volume} {113}},\ \bibinfo
  {pages} {163603} (\bibinfo {year} {2014})}\BibitemShut {NoStop}%
\bibitem [{\citenamefont {Bettles}\ \emph {et~al.}(2015)\citenamefont
  {Bettles}, \citenamefont {Gardiner},\ and\ \citenamefont
  {Adams}}]{Bettles2015d}%
  \BibitemOpen
  \bibfield  {author} {\bibinfo {author} {\bibfnamefont {R.~J.}\ \bibnamefont
  {Bettles}}, \bibinfo {author} {\bibfnamefont {S.~A.}\ \bibnamefont
  {Gardiner}}, \ and\ \bibinfo {author} {\bibfnamefont {C.~S.}\ \bibnamefont
  {Adams}},\ }\href {\doibase 10.1103/PhysRevA.92.063822} {\bibfield  {journal}
  {\bibinfo  {journal} {Phys. Rev. A}\ }\textbf {\bibinfo {volume} {92}},\
  \bibinfo {pages} {063822} (\bibinfo {year} {2015})}\BibitemShut {NoStop}%
\bibitem [{\citenamefont {Bettles}\ \emph {et~al.}(2016)\citenamefont
  {Bettles}, \citenamefont {Gardiner},\ and\ \citenamefont
  {Adams}}]{Bettles2016}%
  \BibitemOpen
  \bibfield  {author} {\bibinfo {author} {\bibfnamefont {R.~J.}\ \bibnamefont
  {Bettles}}, \bibinfo {author} {\bibfnamefont {S.~A.}\ \bibnamefont
  {Gardiner}}, \ and\ \bibinfo {author} {\bibfnamefont {C.~S.}\ \bibnamefont
  {Adams}},\ }\href {\doibase 10.1103/PhysRevLett.116.103602} {\bibfield
  {journal} {\bibinfo  {journal} {Phys. Rev. Lett.}\ }\textbf {\bibinfo
  {volume} {116}},\ \bibinfo {pages} {103602} (\bibinfo {year}
  {2016})}\BibitemShut {NoStop}%
\bibitem [{\citenamefont {Chomaz}\ \emph {et~al.}(2012)\citenamefont {Chomaz},
  \citenamefont {Corman}, \citenamefont {Yefsah}, \citenamefont {Desbuquois},\
  and\ \citenamefont {Dalibard}}]{Chomaz2012}%
  \BibitemOpen
  \bibfield  {author} {\bibinfo {author} {\bibfnamefont {L.}~\bibnamefont
  {Chomaz}}, \bibinfo {author} {\bibfnamefont {L.}~\bibnamefont {Corman}},
  \bibinfo {author} {\bibfnamefont {T.}~\bibnamefont {Yefsah}}, \bibinfo
  {author} {\bibfnamefont {R.}~\bibnamefont {Desbuquois}}, \ and\ \bibinfo
  {author} {\bibfnamefont {J.}~\bibnamefont {Dalibard}},\ }\href {\doibase
  10.1088/1367-2630/14/5/055001} {\bibfield  {journal} {\bibinfo  {journal}
  {New J. Phys.}\ }\textbf {\bibinfo {volume} {14}},\ \bibinfo {pages} {055001}
  (\bibinfo {year} {2012})}\BibitemShut {NoStop}%
\bibitem [{\citenamefont {Pellegrino}\ \emph {et~al.}(2014)\citenamefont
  {Pellegrino}, \citenamefont {Bourgain}, \citenamefont {Jennewein},
  \citenamefont {Sortais}, \citenamefont {Browaeys}, \citenamefont {Jenkins},\
  and\ \citenamefont {Ruostekoski}}]{Pellegrino2014a}%
  \BibitemOpen
  \bibfield  {author} {\bibinfo {author} {\bibfnamefont {J.}~\bibnamefont
  {Pellegrino}}, \bibinfo {author} {\bibfnamefont {R.}~\bibnamefont
  {Bourgain}}, \bibinfo {author} {\bibfnamefont {S.}~\bibnamefont {Jennewein}},
  \bibinfo {author} {\bibfnamefont {Y.~R.~P.}\ \bibnamefont {Sortais}},
  \bibinfo {author} {\bibfnamefont {A.}~\bibnamefont {Browaeys}}, \bibinfo
  {author} {\bibfnamefont {S.~D.}\ \bibnamefont {Jenkins}}, \ and\ \bibinfo
  {author} {\bibfnamefont {J.}~\bibnamefont {Ruostekoski}},\ }\href {\doibase
  10.1103/PhysRevLett.113.133602} {\bibfield  {journal} {\bibinfo  {journal}
  {Phys. Rev. Lett.}\ }\textbf {\bibinfo {volume} {113}},\ \bibinfo {pages}
  {133602} (\bibinfo {year} {2014})}\BibitemShut {NoStop}%
\bibitem [{\citenamefont {Kemp}\ \emph {et~al.}()\citenamefont {Kemp},
  \citenamefont {Roof}, \citenamefont {Havey}, \citenamefont {Sokolov},\ and\
  \citenamefont {Kupriyanov}}]{Kemp}%
  \BibitemOpen
  \bibfield  {author} {\bibinfo {author} {\bibfnamefont {K.}~\bibnamefont
  {Kemp}}, \bibinfo {author} {\bibfnamefont {S.~J.}\ \bibnamefont {Roof}},
  \bibinfo {author} {\bibfnamefont {M.~D.}\ \bibnamefont {Havey}}, \bibinfo
  {author} {\bibfnamefont {I.~M.}\ \bibnamefont {Sokolov}}, \ and\ \bibinfo
  {author} {\bibfnamefont {D.~V.}\ \bibnamefont {Kupriyanov}},\ }\href
  {http://arxiv.org/abs/1410.2497} {\ }\Eprint {http://arxiv.org/abs/1410.2497}
  {arXiv:1410.2497} \BibitemShut {NoStop}%
\bibitem [{\citenamefont {Bromley}\ \emph {et~al.}(2016)\citenamefont
  {Bromley}, \citenamefont {Zhu}, \citenamefont {Bishof}, \citenamefont
  {Zhang}, \citenamefont {Bothwell}, \citenamefont {Schachenmayer},
  \citenamefont {Nicholson}, \citenamefont {Kaiser}, \citenamefont {Yelin},
  \citenamefont {Lukin}, \citenamefont {Rey},\ and\ \citenamefont
  {Ye}}]{Bromley2016}%
  \BibitemOpen
  \bibfield  {author} {\bibinfo {author} {\bibfnamefont {S.~L.}\ \bibnamefont
  {Bromley}}, \bibinfo {author} {\bibfnamefont {B.}~\bibnamefont {Zhu}},
  \bibinfo {author} {\bibfnamefont {M.}~\bibnamefont {Bishof}}, \bibinfo
  {author} {\bibfnamefont {X.}~\bibnamefont {Zhang}}, \bibinfo {author}
  {\bibfnamefont {T.}~\bibnamefont {Bothwell}}, \bibinfo {author}
  {\bibfnamefont {J.}~\bibnamefont {Schachenmayer}}, \bibinfo {author}
  {\bibfnamefont {T.~L.}\ \bibnamefont {Nicholson}}, \bibinfo {author}
  {\bibfnamefont {R.}~\bibnamefont {Kaiser}}, \bibinfo {author} {\bibfnamefont
  {S.~F.}\ \bibnamefont {Yelin}}, \bibinfo {author} {\bibfnamefont {M.~D.}\
  \bibnamefont {Lukin}}, \bibinfo {author} {\bibfnamefont {A.~M.}\ \bibnamefont
  {Rey}}, \ and\ \bibinfo {author} {\bibfnamefont {J.}~\bibnamefont {Ye}},\
  }\href {\doibase 10.1038/ncomms11039} {\bibfield  {journal} {\bibinfo
  {journal} {Nat. Commun.}\ }\textbf {\bibinfo {volume} {7}},\ \bibinfo {pages}
  {11039} (\bibinfo {year} {2016})}\BibitemShut {NoStop}%
\bibitem [{\citenamefont {Casabone}\ \emph {et~al.}(2015)\citenamefont
  {Casabone}, \citenamefont {Friebe}, \citenamefont {Brandst{\"{a}}tter},
  \citenamefont {Sch{\"{u}}ppert}, \citenamefont {Blatt},\ and\ \citenamefont
  {Northup}}]{Casabone2015}%
  \BibitemOpen
  \bibfield  {author} {\bibinfo {author} {\bibfnamefont {B.}~\bibnamefont
  {Casabone}}, \bibinfo {author} {\bibfnamefont {K.}~\bibnamefont {Friebe}},
  \bibinfo {author} {\bibfnamefont {B.}~\bibnamefont {Brandst{\"{a}}tter}},
  \bibinfo {author} {\bibfnamefont {K.}~\bibnamefont {Sch{\"{u}}ppert}},
  \bibinfo {author} {\bibfnamefont {R.}~\bibnamefont {Blatt}}, \ and\ \bibinfo
  {author} {\bibfnamefont {T.~E.}\ \bibnamefont {Northup}},\ }\href {\doibase
  10.1103/PhysRevLett.114.023602} {\bibfield  {journal} {\bibinfo  {journal}
  {Phys. Rev. Lett.}\ }\textbf {\bibinfo {volume} {114}},\ \bibinfo {pages}
  {023602} (\bibinfo {year} {2015})}\BibitemShut {NoStop}%
\bibitem [{\citenamefont {Adamo}\ \emph {et~al.}(2012)\citenamefont {Adamo},
  \citenamefont {Ou}, \citenamefont {So}, \citenamefont {Jenkins},
  \citenamefont {{De Angelis}}, \citenamefont {MacDonald}, \citenamefont {{Di
  Fabrizio}}, \citenamefont {Ruostekoski},\ and\ \citenamefont
  {Zheludev}}]{Adamo2012}%
  \BibitemOpen
  \bibfield  {author} {\bibinfo {author} {\bibfnamefont {G.}~\bibnamefont
  {Adamo}}, \bibinfo {author} {\bibfnamefont {J.~Y.}\ \bibnamefont {Ou}},
  \bibinfo {author} {\bibfnamefont {J.~K.}\ \bibnamefont {So}}, \bibinfo
  {author} {\bibfnamefont {S.~D.}\ \bibnamefont {Jenkins}}, \bibinfo {author}
  {\bibfnamefont {F.}~\bibnamefont {{De Angelis}}}, \bibinfo {author}
  {\bibfnamefont {K.~F.}\ \bibnamefont {MacDonald}}, \bibinfo {author}
  {\bibfnamefont {E.}~\bibnamefont {{Di Fabrizio}}}, \bibinfo {author}
  {\bibfnamefont {J.}~\bibnamefont {Ruostekoski}}, \ and\ \bibinfo {author}
  {\bibfnamefont {N.~I.}\ \bibnamefont {Zheludev}},\ }\href {\doibase
  10.1103/PhysRevLett.109.217401} {\bibfield  {journal} {\bibinfo  {journal}
  {Phys. Rev. Lett.}\ }\textbf {\bibinfo {volume} {109}},\ \bibinfo {pages}
  {217401} (\bibinfo {year} {2012})}\BibitemShut {NoStop}%
\bibitem [{\citenamefont {Ross}\ \emph {et~al.}(2016)\citenamefont {Ross},
  \citenamefont {Mirkin},\ and\ \citenamefont {Schatz}}]{Ross2016a}%
  \BibitemOpen
  \bibfield  {author} {\bibinfo {author} {\bibfnamefont {M.~B.}\ \bibnamefont
  {Ross}}, \bibinfo {author} {\bibfnamefont {C.~A.}\ \bibnamefont {Mirkin}}, \
  and\ \bibinfo {author} {\bibfnamefont {G.~C.}\ \bibnamefont {Schatz}},\
  }\href {\doibase 10.1021/acs.jpcc.5b10800} {\bibfield  {journal} {\bibinfo
  {journal} {J. Phys. Chem. C}\ }\textbf {\bibinfo {volume} {120}},\ \bibinfo
  {pages} {816} (\bibinfo {year} {2016})}\BibitemShut {NoStop}%
\bibitem [{\citenamefont {Kr{\"{a}}mer}\ \emph {et~al.}(2016)\citenamefont
  {Kr{\"{a}}mer}, \citenamefont {Ostermann},\ and\ \citenamefont
  {Ritsch}}]{Kramer}%
  \BibitemOpen
  \bibfield  {author} {\bibinfo {author} {\bibfnamefont {S.}~\bibnamefont
  {Kr{\"{a}}mer}}, \bibinfo {author} {\bibfnamefont {L.}~\bibnamefont
  {Ostermann}}, \ and\ \bibinfo {author} {\bibfnamefont {H.}~\bibnamefont
  {Ritsch}},\ }\href {\doibase 10.1209/0295-5075/114/14003} {\bibfield
  {journal} {\bibinfo  {journal} {Europhys. Lett.}\ }\textbf {\bibinfo {volume}
  {114}},\ \bibinfo {pages} {14003} (\bibinfo {year} {2016})}\BibitemShut
  {NoStop}%
\bibitem [{\citenamefont {Olmos}\ \emph {et~al.}(2013)\citenamefont {Olmos},
  \citenamefont {Yu}, \citenamefont {Singh}, \citenamefont {Schreck},
  \citenamefont {Bongs},\ and\ \citenamefont {Lesanovsky}}]{Olmos2013}%
  \BibitemOpen
  \bibfield  {author} {\bibinfo {author} {\bibfnamefont {B.}~\bibnamefont
  {Olmos}}, \bibinfo {author} {\bibfnamefont {D.}~\bibnamefont {Yu}}, \bibinfo
  {author} {\bibfnamefont {Y.}~\bibnamefont {Singh}}, \bibinfo {author}
  {\bibfnamefont {F.}~\bibnamefont {Schreck}}, \bibinfo {author} {\bibfnamefont
  {K.}~\bibnamefont {Bongs}}, \ and\ \bibinfo {author} {\bibfnamefont
  {I.}~\bibnamefont {Lesanovsky}},\ }\href {\doibase
  10.1103/PhysRevLett.110.143602} {\bibfield  {journal} {\bibinfo  {journal}
  {Phys. Rev. Lett.}\ }\textbf {\bibinfo {volume} {110}},\ \bibinfo {pages}
  {143602} (\bibinfo {year} {2013})}\BibitemShut {NoStop}%
\bibitem [{Note1()}]{Note1}%
  \BibitemOpen
  \bibinfo {note} {Cooperativity can also occur from incoherent sources,
  provided the scatterers are indistinguishable to the detectors \cite
  {Oppel2014}}\BibitemShut {NoStop}%
\bibitem [{\citenamefont {Jenkins}\ and\ \citenamefont
  {Ruostekoski}(2012{\natexlab{a}})}]{Jenkins2012}%
  \BibitemOpen
  \bibfield  {author} {\bibinfo {author} {\bibfnamefont {S.~D.}\ \bibnamefont
  {Jenkins}}\ and\ \bibinfo {author} {\bibfnamefont {J.}~\bibnamefont
  {Ruostekoski}},\ }\href {\doibase 10.1103/PhysRevA.86.031602} {\bibfield
  {journal} {\bibinfo  {journal} {Phys. Rev. A}\ }\textbf {\bibinfo {volume}
  {86}},\ \bibinfo {pages} {031602} (\bibinfo {year}
  {2012}{\natexlab{a}})}\BibitemShut {NoStop}%
\bibitem [{\citenamefont {Nienhuis}\ and\ \citenamefont
  {Schuller}(1987)}]{Nienhuis1987}%
  \BibitemOpen
  \bibfield  {author} {\bibinfo {author} {\bibfnamefont {G.}~\bibnamefont
  {Nienhuis}}\ and\ \bibinfo {author} {\bibfnamefont {F.}~\bibnamefont
  {Schuller}},\ }\href {http://iopscience.iop.org/0022-3700/20/1/008}
  {\bibfield  {journal} {\bibinfo  {journal} {J. Phys. B At. Mol. Phys.}\
  }\textbf {\bibinfo {volume} {20}},\ \bibinfo {pages} {23} (\bibinfo {year}
  {1987})}\BibitemShut {NoStop}%
\bibitem [{\citenamefont {Chang}\ \emph {et~al.}(2012)\citenamefont {Chang},
  \citenamefont {Jiang}, \citenamefont {Gorshkov},\ and\ \citenamefont
  {Kimble}}]{Chang2012}%
  \BibitemOpen
  \bibfield  {author} {\bibinfo {author} {\bibfnamefont {D.~E.}\ \bibnamefont
  {Chang}}, \bibinfo {author} {\bibfnamefont {L.}~\bibnamefont {Jiang}},
  \bibinfo {author} {\bibfnamefont {A.~V.}\ \bibnamefont {Gorshkov}}, \ and\
  \bibinfo {author} {\bibfnamefont {H.~J.}\ \bibnamefont {Kimble}},\ }\href
  {\doibase 10.1088/1367-2630/14/6/063003} {\bibfield  {journal} {\bibinfo
  {journal} {New J. Phys.}\ }\textbf {\bibinfo {volume} {14}},\ \bibinfo
  {pages} {063003} (\bibinfo {year} {2012})}\BibitemShut {NoStop}%
\bibitem [{\citenamefont {Kr{\"{a}}mer}\ and\ \citenamefont
  {Ritsch}(2015)}]{Kramer2015}%
  \BibitemOpen
  \bibfield  {author} {\bibinfo {author} {\bibfnamefont {S.}~\bibnamefont
  {Kr{\"{a}}mer}}\ and\ \bibinfo {author} {\bibfnamefont {H.}~\bibnamefont
  {Ritsch}},\ }\href {\doibase 10.1140/epjd/e2015-60266-5} {\bibfield
  {journal} {\bibinfo  {journal} {Eur. Phys. J. D}\ }\textbf {\bibinfo {volume}
  {69}},\ \bibinfo {pages} {282} (\bibinfo {year} {2015})}\BibitemShut
  {NoStop}%
\bibitem [{\citenamefont {Sutherland}\ and\ \citenamefont
  {Robicheaux}(2016)}]{Sutherland2016}%
  \BibitemOpen
  \bibfield  {author} {\bibinfo {author} {\bibfnamefont {R.~T.}\ \bibnamefont
  {Sutherland}}\ and\ \bibinfo {author} {\bibfnamefont {F.}~\bibnamefont
  {Robicheaux}},\ }\href {\doibase 10.1103/PhysRevA.94.013847} {\bibfield
  {journal} {\bibinfo  {journal} {Phys. Rev. A}\ }\textbf {\bibinfo {volume}
  {94}},\ \bibinfo {pages} {013847} (\bibinfo {year} {2016})}\BibitemShut
  {NoStop}%
\bibitem [{\citenamefont {Yoo}\ and\ \citenamefont {Paik}(2016)}]{Yoo2016a}%
  \BibitemOpen
  \bibfield  {author} {\bibinfo {author} {\bibfnamefont {S.-M.}\ \bibnamefont
  {Yoo}}\ and\ \bibinfo {author} {\bibfnamefont {S.~M.}\ \bibnamefont {Paik}},\
  }\href {\doibase 10.1364/OE.24.002156} {\bibfield  {journal} {\bibinfo
  {journal} {Opt. Express}\ }\textbf {\bibinfo {volume} {24}},\ \bibinfo
  {pages} {2156} (\bibinfo {year} {2016})}\BibitemShut {NoStop}%
\bibitem [{\citenamefont {Yan}\ \emph {et~al.}(2013)\citenamefont {Yan},
  \citenamefont {Moses}, \citenamefont {Gadway}, \citenamefont {Covey},
  \citenamefont {Hazzard}, \citenamefont {Rey}, \citenamefont {Jin},\ and\
  \citenamefont {Ye}}]{Yan2013}%
  \BibitemOpen
  \bibfield  {author} {\bibinfo {author} {\bibfnamefont {B.}~\bibnamefont
  {Yan}}, \bibinfo {author} {\bibfnamefont {S.~A.}\ \bibnamefont {Moses}},
  \bibinfo {author} {\bibfnamefont {B.}~\bibnamefont {Gadway}}, \bibinfo
  {author} {\bibfnamefont {J.~P.}\ \bibnamefont {Covey}}, \bibinfo {author}
  {\bibfnamefont {K.~R.~A.}\ \bibnamefont {Hazzard}}, \bibinfo {author}
  {\bibfnamefont {A.~M.}\ \bibnamefont {Rey}}, \bibinfo {author} {\bibfnamefont
  {D.~S.}\ \bibnamefont {Jin}}, \ and\ \bibinfo {author} {\bibfnamefont
  {J.}~\bibnamefont {Ye}},\ }\href {\doibase 10.1038/nature12483} {\bibfield
  {journal} {\bibinfo  {journal} {Nature}\ }\textbf {\bibinfo {volume} {501}},\
  \bibinfo {pages} {521} (\bibinfo {year} {2013})}\BibitemShut {NoStop}%
\bibitem [{\citenamefont {Barredo}\ \emph {et~al.}(2015)\citenamefont
  {Barredo}, \citenamefont {Labuhn}, \citenamefont {Ravets}, \citenamefont
  {Lahaye}, \citenamefont {Browaeys},\ and\ \citenamefont
  {Adams}}]{Barredo2015}%
  \BibitemOpen
  \bibfield  {author} {\bibinfo {author} {\bibfnamefont {D.}~\bibnamefont
  {Barredo}}, \bibinfo {author} {\bibfnamefont {H.}~\bibnamefont {Labuhn}},
  \bibinfo {author} {\bibfnamefont {S.}~\bibnamefont {Ravets}}, \bibinfo
  {author} {\bibfnamefont {T.}~\bibnamefont {Lahaye}}, \bibinfo {author}
  {\bibfnamefont {A.}~\bibnamefont {Browaeys}}, \ and\ \bibinfo {author}
  {\bibfnamefont {C.~S.}\ \bibnamefont {Adams}},\ }\href {\doibase
  10.1103/PhysRevLett.114.113002} {\bibfield  {journal} {\bibinfo  {journal}
  {Phys. Rev. Lett.}\ }\textbf {\bibinfo {volume} {114}},\ \bibinfo {pages}
  {113002} (\bibinfo {year} {2015})}\BibitemShut {NoStop}%
\bibitem [{\citenamefont {Haakh}\ \emph {et~al.}()\citenamefont {Haakh},
  \citenamefont {Faez},\ and\ \citenamefont {Sandoghdar}}]{Haakh2015}%
  \BibitemOpen
  \bibfield  {author} {\bibinfo {author} {\bibfnamefont {H.~R.}\ \bibnamefont
  {Haakh}}, \bibinfo {author} {\bibfnamefont {S.}~\bibnamefont {Faez}}, \ and\
  \bibinfo {author} {\bibfnamefont {V.}~\bibnamefont {Sandoghdar}},\ }\href
  {http://arxiv.org/abs/1510.07979} {\ }\Eprint
  {http://arxiv.org/abs/1510.07979} {arXiv:1510.07979} \BibitemShut {NoStop}%
\bibitem [{\citenamefont {Liao}\ \emph {et~al.}(2015)\citenamefont {Liao},
  \citenamefont {Zeng}, \citenamefont {Zhu},\ and\ \citenamefont
  {Zubairy}}]{Liao2015}%
  \BibitemOpen
  \bibfield  {author} {\bibinfo {author} {\bibfnamefont {Z.}~\bibnamefont
  {Liao}}, \bibinfo {author} {\bibfnamefont {X.}~\bibnamefont {Zeng}}, \bibinfo
  {author} {\bibfnamefont {S.-Y.}\ \bibnamefont {Zhu}}, \ and\ \bibinfo
  {author} {\bibfnamefont {M.~S.}\ \bibnamefont {Zubairy}},\ }\href {\doibase
  10.1103/PhysRevA.92.023806} {\bibfield  {journal} {\bibinfo  {journal} {Phys.
  Rev. A}\ }\textbf {\bibinfo {volume} {92}},\ \bibinfo {pages} {023806}
  (\bibinfo {year} {2015})}\BibitemShut {NoStop}%
\bibitem [{\citenamefont {Corzo}\ \emph {et~al.}()\citenamefont {Corzo},
  \citenamefont {Gouraud}, \citenamefont {Chandra}, \citenamefont {Goban},
  \citenamefont {Sheremet}, \citenamefont {Kupriyanov},\ and\ \citenamefont
  {Laurat}}]{Corzo2016}%
  \BibitemOpen
  \bibfield  {author} {\bibinfo {author} {\bibfnamefont {N.~V.}\ \bibnamefont
  {Corzo}}, \bibinfo {author} {\bibfnamefont {B.}~\bibnamefont {Gouraud}},
  \bibinfo {author} {\bibfnamefont {A.}~\bibnamefont {Chandra}}, \bibinfo
  {author} {\bibfnamefont {A.}~\bibnamefont {Goban}}, \bibinfo {author}
  {\bibfnamefont {A.~S.}\ \bibnamefont {Sheremet}}, \bibinfo {author}
  {\bibfnamefont {D.~V.}\ \bibnamefont {Kupriyanov}}, \ and\ \bibinfo {author}
  {\bibfnamefont {J.}~\bibnamefont {Laurat}},\ }\href
  {http://arxiv.org/abs/1604.03129} {\ }\Eprint
  {http://arxiv.org/abs/1604.03129} {arXiv:1604.03129} \BibitemShut {NoStop}%
\bibitem [{\citenamefont {S{\o}rensen}\ \emph {et~al.}()\citenamefont
  {S{\o}rensen}, \citenamefont {B{\'{e}}guin}, \citenamefont {Kluge},
  \citenamefont {Iakoupov}, \citenamefont {S{\o}rensen}, \citenamefont
  {M{\"{u}}ller}, \citenamefont {Polzik},\ and\ \citenamefont
  {Appel}}]{Sorensen2016a}%
  \BibitemOpen
  \bibfield  {author} {\bibinfo {author} {\bibfnamefont {H.~L.}\ \bibnamefont
  {S{\o}rensen}}, \bibinfo {author} {\bibfnamefont {J.~B.}\ \bibnamefont
  {B{\'{e}}guin}}, \bibinfo {author} {\bibfnamefont {K.~W.}\ \bibnamefont
  {Kluge}}, \bibinfo {author} {\bibfnamefont {I.}~\bibnamefont {Iakoupov}},
  \bibinfo {author} {\bibfnamefont {A.~S.}\ \bibnamefont {S{\o}rensen}},
  \bibinfo {author} {\bibfnamefont {J.~H.}\ \bibnamefont {M{\"{u}}ller}},
  \bibinfo {author} {\bibfnamefont {E.~S.}\ \bibnamefont {Polzik}}, \ and\
  \bibinfo {author} {\bibfnamefont {J.}~\bibnamefont {Appel}},\ }\href
  {http://arxiv.org/abs/1601.04869} {\ }\Eprint
  {http://arxiv.org/abs/1601.04869} {arXiv:1601.04869} \BibitemShut {NoStop}%
\bibitem [{\citenamefont {Jen}\ \emph {et~al.}(2016)\citenamefont {Jen},
  \citenamefont {Chang},\ and\ \citenamefont {Chen}}]{Jen2016}%
  \BibitemOpen
  \bibfield  {author} {\bibinfo {author} {\bibfnamefont {H.~H.}\ \bibnamefont
  {Jen}}, \bibinfo {author} {\bibfnamefont {M.-S.}\ \bibnamefont {Chang}}, \
  and\ \bibinfo {author} {\bibfnamefont {Y.-C.}\ \bibnamefont {Chen}},\ }\href
  {\doibase 10.1103/PhysRevA.94.013803} {\bibfield  {journal} {\bibinfo
  {journal} {Phys. Rev. A}\ }\textbf {\bibinfo {volume} {94}},\ \bibinfo
  {pages} {013803} (\bibinfo {year} {2016})}\BibitemShut {NoStop}%
\bibitem [{\citenamefont {Eschner}\ \emph {et~al.}(2001)\citenamefont
  {Eschner}, \citenamefont {Raab}, \citenamefont {Schmidt-Kaler},\ and\
  \citenamefont {Blatt}}]{Eschner2001}%
  \BibitemOpen
  \bibfield  {author} {\bibinfo {author} {\bibfnamefont {J.}~\bibnamefont
  {Eschner}}, \bibinfo {author} {\bibfnamefont {C.}~\bibnamefont {Raab}},
  \bibinfo {author} {\bibfnamefont {F.}~\bibnamefont {Schmidt-Kaler}}, \ and\
  \bibinfo {author} {\bibfnamefont {R.}~\bibnamefont {Blatt}},\ }\href
  {\doibase 10.1038/35097017} {\bibfield  {journal} {\bibinfo  {journal}
  {Nature}\ }\textbf {\bibinfo {volume} {413}},\ \bibinfo {pages} {495}
  (\bibinfo {year} {2001})}\BibitemShut {NoStop}%
\bibitem [{\citenamefont {Svidzinsky}\ \emph {et~al.}(2013)\citenamefont
  {Svidzinsky}, \citenamefont {Yuan},\ and\ \citenamefont
  {Scully}}]{Svidzinsky2013}%
  \BibitemOpen
  \bibfield  {author} {\bibinfo {author} {\bibfnamefont {A.~A.}\ \bibnamefont
  {Svidzinsky}}, \bibinfo {author} {\bibfnamefont {L.}~\bibnamefont {Yuan}}, \
  and\ \bibinfo {author} {\bibfnamefont {M.~O.}\ \bibnamefont {Scully}},\
  }\href {\doibase 10.1103/PhysRevX.3.041001} {\bibfield  {journal} {\bibinfo
  {journal} {Phys. Rev. X}\ }\textbf {\bibinfo {volume} {3}},\ \bibinfo {pages}
  {041001} (\bibinfo {year} {2013})}\BibitemShut {NoStop}%
\bibitem [{\citenamefont {Zhou}\ \emph {et~al.}(2010)\citenamefont {Zhou},
  \citenamefont {Xu}, \citenamefont {Chen},\ and\ \citenamefont
  {Chen}}]{Zhou2010}%
  \BibitemOpen
  \bibfield  {author} {\bibinfo {author} {\bibfnamefont {X.}~\bibnamefont
  {Zhou}}, \bibinfo {author} {\bibfnamefont {X.}~\bibnamefont {Xu}}, \bibinfo
  {author} {\bibfnamefont {X.}~\bibnamefont {Chen}}, \ and\ \bibinfo {author}
  {\bibfnamefont {J.}~\bibnamefont {Chen}},\ }\href {\doibase
  10.1103/PhysRevA.81.012115} {\bibfield  {journal} {\bibinfo  {journal} {Phys.
  Rev. A}\ }\textbf {\bibinfo {volume} {81}},\ \bibinfo {pages} {012115}
  (\bibinfo {year} {2010})}\BibitemShut {NoStop}%
\bibitem [{\citenamefont {Barber}\ \emph {et~al.}(2008)\citenamefont {Barber},
  \citenamefont {Stalnaker}, \citenamefont {Lemke}, \citenamefont {Poli},
  \citenamefont {Oates}, \citenamefont {Fortier}, \citenamefont {Diddams},
  \citenamefont {Hollberg}, \citenamefont {Hoyt}, \citenamefont
  {Taichenachev},\ and\ \citenamefont {Yudin}}]{Barber2008}%
  \BibitemOpen
  \bibfield  {author} {\bibinfo {author} {\bibfnamefont {Z.~W.}\ \bibnamefont
  {Barber}}, \bibinfo {author} {\bibfnamefont {J.~E.}\ \bibnamefont
  {Stalnaker}}, \bibinfo {author} {\bibfnamefont {N.~D.}\ \bibnamefont
  {Lemke}}, \bibinfo {author} {\bibfnamefont {N.}~\bibnamefont {Poli}},
  \bibinfo {author} {\bibfnamefont {C.~W.}\ \bibnamefont {Oates}}, \bibinfo
  {author} {\bibfnamefont {T.~M.}\ \bibnamefont {Fortier}}, \bibinfo {author}
  {\bibfnamefont {S.~A.}\ \bibnamefont {Diddams}}, \bibinfo {author}
  {\bibfnamefont {L.}~\bibnamefont {Hollberg}}, \bibinfo {author}
  {\bibfnamefont {C.~W.}\ \bibnamefont {Hoyt}}, \bibinfo {author}
  {\bibfnamefont {A.~V.}\ \bibnamefont {Taichenachev}}, \ and\ \bibinfo
  {author} {\bibfnamefont {V.~I.}\ \bibnamefont {Yudin}},\ }\href {\doibase
  10.1103/PhysRevLett.100.103002} {\bibfield  {journal} {\bibinfo  {journal}
  {Phys. Rev. Lett.}\ }\textbf {\bibinfo {volume} {100}},\ \bibinfo {pages}
  {103002} (\bibinfo {year} {2008})}\BibitemShut {NoStop}%
\bibitem [{\citenamefont {Fukuhara}\ \emph {et~al.}(2009)\citenamefont
  {Fukuhara}, \citenamefont {Sugawa}, \citenamefont {Sugimoto}, \citenamefont
  {Taie},\ and\ \citenamefont {Takahashi}}]{Fukuhara2009a}%
  \BibitemOpen
  \bibfield  {author} {\bibinfo {author} {\bibfnamefont {T.}~\bibnamefont
  {Fukuhara}}, \bibinfo {author} {\bibfnamefont {S.}~\bibnamefont {Sugawa}},
  \bibinfo {author} {\bibfnamefont {M.}~\bibnamefont {Sugimoto}}, \bibinfo
  {author} {\bibfnamefont {S.}~\bibnamefont {Taie}}, \ and\ \bibinfo {author}
  {\bibfnamefont {Y.}~\bibnamefont {Takahashi}},\ }\href {\doibase
  10.1103/PhysRevA.79.041604} {\bibfield  {journal} {\bibinfo  {journal} {Phys.
  Rev. A}\ }\textbf {\bibinfo {volume} {79}},\ \bibinfo {pages} {041604}
  (\bibinfo {year} {2009})}\BibitemShut {NoStop}%
\bibitem [{\citenamefont {Stellmer}\ \emph {et~al.}(2012)\citenamefont
  {Stellmer}, \citenamefont {Pasquiou}, \citenamefont {Grimm},\ and\
  \citenamefont {Schreck}}]{Stellmer2012}%
  \BibitemOpen
  \bibfield  {author} {\bibinfo {author} {\bibfnamefont {S.}~\bibnamefont
  {Stellmer}}, \bibinfo {author} {\bibfnamefont {B.}~\bibnamefont {Pasquiou}},
  \bibinfo {author} {\bibfnamefont {R.}~\bibnamefont {Grimm}}, \ and\ \bibinfo
  {author} {\bibfnamefont {F.}~\bibnamefont {Schreck}},\ }\href {\doibase
  10.1103/PhysRevLett.109.115302} {\bibfield  {journal} {\bibinfo  {journal}
  {Phys. Rev. Lett.}\ }\textbf {\bibinfo {volume} {109}},\ \bibinfo {pages}
  {115302} (\bibinfo {year} {2012})}\BibitemShut {NoStop}%
\bibitem [{\citenamefont {Nogrette}\ \emph {et~al.}(2014)\citenamefont
  {Nogrette}, \citenamefont {Labuhn}, \citenamefont {Ravets}, \citenamefont
  {Barredo}, \citenamefont {B{\'{e}}guin}, \citenamefont {Vernier},
  \citenamefont {Lahaye},\ and\ \citenamefont {Browaeys}}]{Nogrette2014}%
  \BibitemOpen
  \bibfield  {author} {\bibinfo {author} {\bibfnamefont {F.}~\bibnamefont
  {Nogrette}}, \bibinfo {author} {\bibfnamefont {H.}~\bibnamefont {Labuhn}},
  \bibinfo {author} {\bibfnamefont {S.}~\bibnamefont {Ravets}}, \bibinfo
  {author} {\bibfnamefont {D.}~\bibnamefont {Barredo}}, \bibinfo {author}
  {\bibfnamefont {L.}~\bibnamefont {B{\'{e}}guin}}, \bibinfo {author}
  {\bibfnamefont {A.}~\bibnamefont {Vernier}}, \bibinfo {author} {\bibfnamefont
  {T.}~\bibnamefont {Lahaye}}, \ and\ \bibinfo {author} {\bibfnamefont
  {A.}~\bibnamefont {Browaeys}},\ }\href {\doibase 10.1103/PhysRevX.4.021034}
  {\bibfield  {journal} {\bibinfo  {journal} {Phys. Rev. X}\ }\textbf {\bibinfo
  {volume} {4}},\ \bibinfo {pages} {021034} (\bibinfo {year}
  {2014})}\BibitemShut {NoStop}%
\bibitem [{\citenamefont {Lester}\ \emph {et~al.}(2015)\citenamefont {Lester},
  \citenamefont {Luick}, \citenamefont {Kaufman}, \citenamefont {Reynolds},\
  and\ \citenamefont {Regal}}]{Lester2015a}%
  \BibitemOpen
  \bibfield  {author} {\bibinfo {author} {\bibfnamefont {B.~J.}\ \bibnamefont
  {Lester}}, \bibinfo {author} {\bibfnamefont {N.}~\bibnamefont {Luick}},
  \bibinfo {author} {\bibfnamefont {A.~M.}\ \bibnamefont {Kaufman}}, \bibinfo
  {author} {\bibfnamefont {C.~M.}\ \bibnamefont {Reynolds}}, \ and\ \bibinfo
  {author} {\bibfnamefont {C.~A.}\ \bibnamefont {Regal}},\ }\href {\doibase
  10.1103/PhysRevLett.115.073003} {\bibfield  {journal} {\bibinfo  {journal}
  {Phys. Rev. Lett.}\ }\textbf {\bibinfo {volume} {115}},\ \bibinfo {pages}
  {073003} (\bibinfo {year} {2015})}\BibitemShut {NoStop}%
\bibitem [{Note2()}]{Note2}%
  \BibitemOpen
  \bibinfo {note} {We assume $\omega \gg \gamma _0$ and so for all detunings in
  this paper, ${\lambda \simeq \lambda _0=2\pi c/\omega _0}$.}\BibitemShut
  {Stop}%
\bibitem [{\citenamefont {Gardiner}\ and\ \citenamefont
  {Zoller}(2015)}]{GardinerZoller2}%
  \BibitemOpen
  \bibfield  {author} {\bibinfo {author} {\bibfnamefont {C.}~\bibnamefont
  {Gardiner}}\ and\ \bibinfo {author} {\bibfnamefont {P.}~\bibnamefont
  {Zoller}},\ }\href@noop {} {\emph {\bibinfo {title} {{The Quantum World of
  Ultra-Cold Atoms and Light: Book 2 The Physics of Quantum-Optical
  Devices}}}},\ \bibinfo {edition} {1st}\ ed.\ (\bibinfo  {publisher} {Imperial
  College Press},\ \bibinfo {address} {London},\ \bibinfo {year}
  {2015})\BibitemShut {NoStop}%
\bibitem [{\citenamefont {Javanainen}\ \emph {et~al.}(1999)\citenamefont
  {Javanainen}, \citenamefont {Ruostekoski}, \citenamefont {Vestergaard},\ and\
  \citenamefont {Francis}}]{Javanainen1999}%
  \BibitemOpen
  \bibfield  {author} {\bibinfo {author} {\bibfnamefont {J.}~\bibnamefont
  {Javanainen}}, \bibinfo {author} {\bibfnamefont {J.}~\bibnamefont
  {Ruostekoski}}, \bibinfo {author} {\bibfnamefont {B.}~\bibnamefont
  {Vestergaard}}, \ and\ \bibinfo {author} {\bibfnamefont {M.~R.}\ \bibnamefont
  {Francis}},\ }\href {\doibase 10.1103/PhysRevA.59.649} {\bibfield  {journal}
  {\bibinfo  {journal} {Phys. Rev. A}\ }\textbf {\bibinfo {volume} {59}},\
  \bibinfo {pages} {649} (\bibinfo {year} {1999})}\BibitemShut {NoStop}%
\bibitem [{\citenamefont {Svidzinsky}\ \emph {et~al.}(2010)\citenamefont
  {Svidzinsky}, \citenamefont {Chang},\ and\ \citenamefont
  {Scully}}]{Svidzinsky2010}%
  \BibitemOpen
  \bibfield  {author} {\bibinfo {author} {\bibfnamefont {A.~A.}\ \bibnamefont
  {Svidzinsky}}, \bibinfo {author} {\bibfnamefont {J.~T.}\ \bibnamefont
  {Chang}}, \ and\ \bibinfo {author} {\bibfnamefont {M.~O.}\ \bibnamefont
  {Scully}},\ }\href {\doibase 10.1103/PhysRevA.81.053821} {\bibfield
  {journal} {\bibinfo  {journal} {Phys. Rev. A}\ }\textbf {\bibinfo {volume}
  {81}},\ \bibinfo {pages} {053821} (\bibinfo {year} {2010})}\BibitemShut
  {NoStop}%
\bibitem [{\citenamefont {Bettles}(2016)}]{BettlesThesis}%
  \BibitemOpen
  \bibfield  {author} {\bibinfo {author} {\bibfnamefont {R.~J.}\ \bibnamefont
  {Bettles}},\ }\emph {\bibinfo {title} {{Cooperative Interactions in Lattices
  of Atomic Dipoles}}},\ \href {http://etheses.dur.ac.uk/11636/} {Ph.D.
  thesis},\ \bibinfo  {school} {Durham University} (\bibinfo {year}
  {2016})\BibitemShut {NoStop}%
\bibitem [{\citenamefont {Jackson}(1963)}]{Jackson1963}%
  \BibitemOpen
  \bibfield  {author} {\bibinfo {author} {\bibfnamefont {J.~D.}\ \bibnamefont
  {Jackson}},\ }\href@noop {} {\emph {\bibinfo {title} {{Classical
  Electrodynamics}}}}\ (\bibinfo  {publisher} {John Wiley {\&} Sons, Inc.},\
  \bibinfo {address} {New York, London},\ \bibinfo {year} {1963})\BibitemShut
  {NoStop}%
\bibitem [{\citenamefont {Jenkins}\ and\ \citenamefont
  {Ruostekoski}(2012{\natexlab{b}})}]{Jenkins2012a}%
  \BibitemOpen
  \bibfield  {author} {\bibinfo {author} {\bibfnamefont {S.~D.}\ \bibnamefont
  {Jenkins}}\ and\ \bibinfo {author} {\bibfnamefont {J.}~\bibnamefont
  {Ruostekoski}},\ }\href {\doibase 10.1103/PhysRevB.86.085116} {\bibfield
  {journal} {\bibinfo  {journal} {Phys. Rev. B}\ }\textbf {\bibinfo {volume}
  {86}},\ \bibinfo {pages} {085116} (\bibinfo {year}
  {2012}{\natexlab{b}})}\BibitemShut {NoStop}%
\bibitem [{\citenamefont {Morice}\ \emph {et~al.}(1995)\citenamefont {Morice},
  \citenamefont {Castin},\ and\ \citenamefont {Dalibard}}]{Morice1995}%
  \BibitemOpen
  \bibfield  {author} {\bibinfo {author} {\bibfnamefont {O.}~\bibnamefont
  {Morice}}, \bibinfo {author} {\bibfnamefont {Y.}~\bibnamefont {Castin}}, \
  and\ \bibinfo {author} {\bibfnamefont {J.}~\bibnamefont {Dalibard}},\ }\href
  {\doibase 10.1103/PhysRevA.51.3896} {\bibfield  {journal} {\bibinfo
  {journal} {Phys. Rev. A}\ }\textbf {\bibinfo {volume} {51}},\ \bibinfo
  {pages} {3896} (\bibinfo {year} {1995})}\BibitemShut {NoStop}%
\bibitem [{\citenamefont {Ruostekoski}\ and\ \citenamefont
  {Javanainen}(1997)}]{Ruostekoski1997a}%
  \BibitemOpen
  \bibfield  {author} {\bibinfo {author} {\bibfnamefont {J.}~\bibnamefont
  {Ruostekoski}}\ and\ \bibinfo {author} {\bibfnamefont {J.}~\bibnamefont
  {Javanainen}},\ }\href {\doibase 10.1103/PhysRevA.55.513} {\bibfield
  {journal} {\bibinfo  {journal} {Phys. Rev. A}\ }\textbf {\bibinfo {volume}
  {55}},\ \bibinfo {pages} {513} (\bibinfo {year} {1997})}\BibitemShut
  {NoStop}%
\bibitem [{\citenamefont {Samoylova}\ \emph {et~al.}(2014)\citenamefont
  {Samoylova}, \citenamefont {Piovella}, \citenamefont {Bachelard},\ and\
  \citenamefont {Courteille}}]{Samoylova2014}%
  \BibitemOpen
  \bibfield  {author} {\bibinfo {author} {\bibfnamefont {M.}~\bibnamefont
  {Samoylova}}, \bibinfo {author} {\bibfnamefont {N.}~\bibnamefont {Piovella}},
  \bibinfo {author} {\bibfnamefont {R.}~\bibnamefont {Bachelard}}, \ and\
  \bibinfo {author} {\bibfnamefont {P.~W.}\ \bibnamefont {Courteille}},\ }\href
  {\doibase 10.1016/j.optcom.2013.09.016} {\bibfield  {journal} {\bibinfo
  {journal} {Opt. Commun.}\ }\textbf {\bibinfo {volume} {312}},\ \bibinfo
  {pages} {94} (\bibinfo {year} {2014})}\BibitemShut {NoStop}%
\bibitem [{\citenamefont {Weller}\ \emph {et~al.}(2011)\citenamefont {Weller},
  \citenamefont {Bettles}, \citenamefont {Siddons}, \citenamefont {Adams},\
  and\ \citenamefont {Hughes}}]{Weller2011}%
  \BibitemOpen
  \bibfield  {author} {\bibinfo {author} {\bibfnamefont {L.}~\bibnamefont
  {Weller}}, \bibinfo {author} {\bibfnamefont {R.~J.}\ \bibnamefont {Bettles}},
  \bibinfo {author} {\bibfnamefont {P.}~\bibnamefont {Siddons}}, \bibinfo
  {author} {\bibfnamefont {C.~S.}\ \bibnamefont {Adams}}, \ and\ \bibinfo
  {author} {\bibfnamefont {I.~G.}\ \bibnamefont {Hughes}},\ }\href {\doibase
  10.1088/0953-4075/44/19/195006} {\bibfield  {journal} {\bibinfo  {journal}
  {J. Phys. B At. Mol. Opt. Phys.}\ }\textbf {\bibinfo {volume} {44}},\
  \bibinfo {pages} {195006} (\bibinfo {year} {2011})}\BibitemShut {NoStop}%
\bibitem [{Note3()}]{Note3}%
  \BibitemOpen
  \bibinfo {note} {A discussion of the conditions for when $\protect \mathsf
  {M}$ is and is not invertible are left to later work.}\BibitemShut {Stop}%
\bibitem [{Note4()}]{Note4}%
  \BibitemOpen
  \bibinfo {note} {We define our notation for the dot product of complex column
  vectors as $\protect \mathaccentV {vec}17E{\protect \mathbf {a}}^*\cdot
  \protect \mathaccentV {vec}17E{\protect \mathbf {b}}\equiv \DOTSB \sum@
  \slimits@ _p a_p^* b_p$, where $^*$ denotes a complex conjugate.}\BibitemShut
  {Stop}%
\bibitem [{\citenamefont {Milonni}\ and\ \citenamefont
  {Knight}(1973)}]{Milonni1973}%
  \BibitemOpen
  \bibfield  {author} {\bibinfo {author} {\bibfnamefont {P.~W.}\ \bibnamefont
  {Milonni}}\ and\ \bibinfo {author} {\bibfnamefont {P.~L.}\ \bibnamefont
  {Knight}},\ }\href
  {http://www.sciencedirect.com/science/article/pii/0030401873902393}
  {\bibfield  {journal} {\bibinfo  {journal} {Opt. Commun.}\ }\textbf {\bibinfo
  {volume} {9}},\ \bibinfo {pages} {119} (\bibinfo {year} {1973})}\BibitemShut
  {NoStop}%
\bibitem [{\citenamefont {Feng}\ \emph {et~al.}(2013)\citenamefont {Feng},
  \citenamefont {Li},\ and\ \citenamefont {Zhu}}]{Feng2013a}%
  \BibitemOpen
  \bibfield  {author} {\bibinfo {author} {\bibfnamefont {W.}~\bibnamefont
  {Feng}}, \bibinfo {author} {\bibfnamefont {Y.}~\bibnamefont {Li}}, \ and\
  \bibinfo {author} {\bibfnamefont {S.~Y.}\ \bibnamefont {Zhu}},\ }\href
  {\doibase 10.1103/PhysRevA.88.033856} {\bibfield  {journal} {\bibinfo
  {journal} {Phys. Rev. A}\ }\textbf {\bibinfo {volume} {88}},\ \bibinfo
  {pages} {033856} (\bibinfo {year} {2013})}\BibitemShut {NoStop}%
\bibitem [{Note5()}]{Note5}%
  \BibitemOpen
  \bibinfo {note} {The mode numbering is chosen so as to be compatible with the
  mode numbering in Fig.\ \ref {fig:NAtomsChainModeVectors}. The prime
  indicates modes polarized parallel to the atomic axis.}\BibitemShut {Stop}%
\bibitem [{\citenamefont {Hopkins}\ \emph {et~al.}(2015)\citenamefont
  {Hopkins}, \citenamefont {Filonov}, \citenamefont {Glybovski},\ and\
  \citenamefont {Miroshnichenko}}]{Hopkins2015}%
  \BibitemOpen
  \bibfield  {author} {\bibinfo {author} {\bibfnamefont {B.}~\bibnamefont
  {Hopkins}}, \bibinfo {author} {\bibfnamefont {D.~S.}\ \bibnamefont
  {Filonov}}, \bibinfo {author} {\bibfnamefont {S.~B.}\ \bibnamefont
  {Glybovski}}, \ and\ \bibinfo {author} {\bibfnamefont {A.~E.}\ \bibnamefont
  {Miroshnichenko}},\ }\href {\doibase 10.1103/PhysRevB.92.045433} {\bibfield
  {journal} {\bibinfo  {journal} {Phys. Rev. B}\ }\textbf {\bibinfo {volume}
  {92}},\ \bibinfo {pages} {045433} (\bibinfo {year} {2015})}\BibitemShut
  {NoStop}%
\bibitem [{\citenamefont {Chong}\ \emph {et~al.}(2014)\citenamefont {Chong},
  \citenamefont {Hopkins}, \citenamefont {Staude}, \citenamefont
  {Miroshnichenko}, \citenamefont {Dominguez}, \citenamefont {Decker},
  \citenamefont {Neshev}, \citenamefont {Brener},\ and\ \citenamefont
  {Kivshar}}]{Chong2014}%
  \BibitemOpen
  \bibfield  {author} {\bibinfo {author} {\bibfnamefont {K.~E.}\ \bibnamefont
  {Chong}}, \bibinfo {author} {\bibfnamefont {B.}~\bibnamefont {Hopkins}},
  \bibinfo {author} {\bibfnamefont {I.}~\bibnamefont {Staude}}, \bibinfo
  {author} {\bibfnamefont {A.~E.}\ \bibnamefont {Miroshnichenko}}, \bibinfo
  {author} {\bibfnamefont {J.}~\bibnamefont {Dominguez}}, \bibinfo {author}
  {\bibfnamefont {M.}~\bibnamefont {Decker}}, \bibinfo {author} {\bibfnamefont
  {D.~N.}\ \bibnamefont {Neshev}}, \bibinfo {author} {\bibfnamefont
  {I.}~\bibnamefont {Brener}}, \ and\ \bibinfo {author} {\bibfnamefont {Y.~S.}\
  \bibnamefont {Kivshar}},\ }\href {\doibase 10.1002/smll.201303612} {\bibfield
   {journal} {\bibinfo  {journal} {Small}\ }\textbf {\bibinfo {volume} {10}},\
  \bibinfo {pages} {1985} (\bibinfo {year} {2014})}\BibitemShut {NoStop}%
\bibitem [{\citenamefont {Emami}\ \emph {et~al.}(2015)\citenamefont {Emami},
  \citenamefont {Soltanian}, \citenamefont {Attaran}, \citenamefont
  {Abdul-Rashid}, \citenamefont {Penny}, \citenamefont {Moghavvemi},
  \citenamefont {Harun}, \citenamefont {Ahmad},\ and\ \citenamefont
  {Mohammed}}]{Emami2014}%
  \BibitemOpen
  \bibfield  {author} {\bibinfo {author} {\bibfnamefont {S.~D.}\ \bibnamefont
  {Emami}}, \bibinfo {author} {\bibfnamefont {M.~R.~K.}\ \bibnamefont
  {Soltanian}}, \bibinfo {author} {\bibfnamefont {A.}~\bibnamefont {Attaran}},
  \bibinfo {author} {\bibfnamefont {H.~A.}\ \bibnamefont {Abdul-Rashid}},
  \bibinfo {author} {\bibfnamefont {R.}~\bibnamefont {Penny}}, \bibinfo
  {author} {\bibfnamefont {M.}~\bibnamefont {Moghavvemi}}, \bibinfo {author}
  {\bibfnamefont {S.~W.}\ \bibnamefont {Harun}}, \bibinfo {author}
  {\bibfnamefont {H.}~\bibnamefont {Ahmad}}, \ and\ \bibinfo {author}
  {\bibfnamefont {W.~S.}\ \bibnamefont {Mohammed}},\ }\href {\doibase
  10.1007/s00339-014-8832-2} {\bibfield  {journal} {\bibinfo  {journal} {Appl.
  Phys. A}\ }\textbf {\bibinfo {volume} {118}},\ \bibinfo {pages} {139}
  (\bibinfo {year} {2015})}\BibitemShut {NoStop}%
\bibitem [{\citenamefont {Christofi}\ \emph {et~al.}(2016)\citenamefont
  {Christofi}, \citenamefont {Pinheiro},\ and\ \citenamefont {{Dal
  Negro}}}]{Christofi2016}%
  \BibitemOpen
  \bibfield  {author} {\bibinfo {author} {\bibfnamefont {A.}~\bibnamefont
  {Christofi}}, \bibinfo {author} {\bibfnamefont {F.~A.}\ \bibnamefont
  {Pinheiro}}, \ and\ \bibinfo {author} {\bibfnamefont {L.}~\bibnamefont {{Dal
  Negro}}},\ }\href {\doibase 10.1364/OL.41.001933} {\bibfield  {journal}
  {\bibinfo  {journal} {Opt. Lett.}\ }\textbf {\bibinfo {volume} {41}},\
  \bibinfo {pages} {1933} (\bibinfo {year} {2016})}\BibitemShut {NoStop}%
\bibitem [{\citenamefont {Skipetrov}\ and\ \citenamefont
  {Goetschy}(2011)}]{Skipetrov2011}%
  \BibitemOpen
  \bibfield  {author} {\bibinfo {author} {\bibfnamefont {S.~E.}\ \bibnamefont
  {Skipetrov}}\ and\ \bibinfo {author} {\bibfnamefont {A.}~\bibnamefont
  {Goetschy}},\ }\href {\doibase 10.1088/1751-8113/44/6/065102} {\bibfield
  {journal} {\bibinfo  {journal} {J. Phys. A Math. Theor.}\ }\textbf {\bibinfo
  {volume} {44}},\ \bibinfo {pages} {065102} (\bibinfo {year}
  {2011})}\BibitemShut {NoStop}%
\bibitem [{\citenamefont {Bellando}\ \emph {et~al.}(2014)\citenamefont
  {Bellando}, \citenamefont {Gero}, \citenamefont {Akkermans},\ and\
  \citenamefont {Kaiser}}]{Bellando2014}%
  \BibitemOpen
  \bibfield  {author} {\bibinfo {author} {\bibfnamefont {L.}~\bibnamefont
  {Bellando}}, \bibinfo {author} {\bibfnamefont {A.}~\bibnamefont {Gero}},
  \bibinfo {author} {\bibfnamefont {E.}~\bibnamefont {Akkermans}}, \ and\
  \bibinfo {author} {\bibfnamefont {R.}~\bibnamefont {Kaiser}},\ }\href
  {\doibase 10.1103/PhysRevA.90.063822} {\bibfield  {journal} {\bibinfo
  {journal} {Phys. Rev. A}\ }\textbf {\bibinfo {volume} {90}},\ \bibinfo
  {pages} {063822} (\bibinfo {year} {2014})}\BibitemShut {NoStop}%
\bibitem [{Note6()}]{Note6}%
  \BibitemOpen
  \bibinfo {note} {This form for $\protect \mathbf {d}_j^{\ell }$ is a good
  approximation for some, but not all, of the eigenvectors}\BibitemShut
  {NoStop}%
\bibitem [{\citenamefont {Markel}(1993)}]{Markel1993}%
  \BibitemOpen
  \bibfield  {author} {\bibinfo {author} {\bibfnamefont {V.~A.}\ \bibnamefont
  {Markel}},\ }\href {\doibase 10.1080/09500349314552291} {\bibfield  {journal}
  {\bibinfo  {journal} {J. Mod. Opt.}\ }\textbf {\bibinfo {volume} {40}},\
  \bibinfo {pages} {2281} (\bibinfo {year} {1993})}\BibitemShut {NoStop}%
\bibitem [{Note7()}]{Note7}%
  \BibitemOpen
  \bibinfo {note} {The finite shifts of the black dashed lines in Fig.\ \ref
  {fig:NAtomChainShiftsWidthsInf}(c) at $a=2.5\lambda $ and $a=3.5\lambda $ are
  simply the computational limit rather than any physical limit.}\BibitemShut
  {Stop}%
\bibitem [{\citenamefont {Lee}\ \emph {et~al.}(2016)\citenamefont {Lee},
  \citenamefont {Jenkins},\ and\ \citenamefont {Ruostekoski}}]{Lee2016}%
  \BibitemOpen
  \bibfield  {author} {\bibinfo {author} {\bibfnamefont {M.~D.}\ \bibnamefont
  {Lee}}, \bibinfo {author} {\bibfnamefont {S.~D.}\ \bibnamefont {Jenkins}}, \
  and\ \bibinfo {author} {\bibfnamefont {J.}~\bibnamefont {Ruostekoski}},\
  }\href {\doibase 10.1103/PhysRevA.93.063803} {\bibfield  {journal} {\bibinfo
  {journal} {Phys. Rev. A}\ }\textbf {\bibinfo {volume} {93}},\ \bibinfo
  {pages} {063803} (\bibinfo {year} {2016})}\BibitemShut {NoStop}%
\bibitem [{Note8()}]{Note8}%
  \BibitemOpen
  \bibinfo {note} {DOI: \protect \href
  {http://dx.doi.org/10.15128/r2gx41mh849}{10.15128/r2gx41mh849}}\BibitemShut
  {NoStop}%
\end{thebibliography}%

\end{document}